\documentclass[aps,pre,twocolumn,groupedaddress]{revtex4-2}

\usepackage{tikz}
\usepackage{amsmath}
\usepackage{hyperref}
\usepackage{cleveref}
\usepackage{subcaption}
\usepackage{url}        

\hypersetup{
pdfauthor = {Pietro Sillano, Siewert Jan Marrink, Timon Idema},
pdftitle = {MesoMem: A mesoscale membrane model based on an additive potential},
colorlinks = true,	  
linkcolor = black,        
citecolor = black,        
filecolor = black,         
urlcolor = black
}

\makeatletter
\let\selectlanguage\@gobble
\makeatother

\begin{document}

\title{MesoMem: A mesoscale membrane model based on an additive potential}
\author{Pietro Sillano}
\affiliation{Department of Bionanoscience, Kavli Institute of Nanoscience, Delft University of Technology, Delft, The Netherlands}
\affiliation{Molecular Dynamics Group, Groningen Biomolecular Sciences and Biotechnology Institute, University of Groningen, The Netherlands}

\author{Siewert J. Marrink}
\affiliation{Molecular Dynamics Group, Groningen Biomolecular Sciences and Biotechnology Institute, University of Groningen, The Netherlands}

\author{Timon Idema} \email{t.idema@tudelft.nl}
\affiliation{Department of Bionanoscience, Kavli Institute of Nanoscience, Delft University of Technology, Delft, The Netherlands}

\date{\today}

\begin{abstract}
Bridging the gap between atomistic detail and continuum mechanics is a central challenge in modeling biological membranes, particularly for mesoscopic phenomena spanning large length and time scales. In this work, we introduce a new, solvent-free, one-particle-thick, coarse-grained model for lipid bilayers, governed by an additive potential. Our approach treats orientational elasticity through distinct additive energy terms for tilt and splay, offering an unbiased potential form. The model is implemented in the LAMMPS molecular dynamics engine. Our simulations show spontaneous self-assembly of lamellar structures and stable vesicles from disordered states. We map the dynamical phase diagram of the system, identifying distinct gel-like, fluid, and gas regimes, controlled by temperature and the steepness of the isotropic attraction. The model accurately reproduces the theoretical $1/q^{4}$ fluctuation spectrum for tensionless membranes and exhibits tunable mechanical properties, including biologically relevant bending rigidities. We show how we can include osmotic pressure and spontaneous curvature in our model. Finally, we demonstrate the model's applicability to complex membrane remodeling by simulating the adhesive wrapping of colloidal nanoparticles, recovering the predicted dependency on particle size and adhesion strength.
\end{abstract}

\maketitle

\section{Introduction}

Biological membranes are essential structural elements of living systems. They provide the boundaries of cells and organelles, mediate selective transport, and act as active participants in mechanical and biochemical processes~\cite{albertsMolecularBiologyCell2002}. From a physical perspective, membranes are thin, flexible surfaces with rich collective behavior, encompassing thermal undulations, curvature generation, and large-scale morphological transformations such as budding and fusion~\cite{freyMoreJustBarrier2021}. Capturing this complexity in a computational framework remains a central challenge in soft matter physics. 
The primary difficulty arises from the wide separation of length and time scales that characterize membrane dynamics~\cite{noguchiMembraneSimulationModels2009,marrinkComputationalModelingRealistic2019}. Events at the membrane range from the nanosecond dynamics of individual lipid molecules to the microsecond-to-second timescales of protein diffusion and macroscopic domain formation, and ultimately to the cellular-level processes that happen over minutes or even longer. 

Atomistic molecular dynamics provides an accurate description of interactions, but is computationally limited to relatively small systems and short timescales. At the other end of the spectrum, continuum elasticity models effectively describe large-scale shapes and bending energetics, but lack explicit microscopic degrees of freedom. Bridging these extremes requires coarse-grained strategies that reduce molecular detail while retaining local interactions, enabling simulations of mesoscopic membrane behavior. Coarse-grained force fields such as Martini~\cite{pedersenMartini3Lipidome2025}, the Cooke-Deserno model \cite{cookeTunableGenericModel2005} or the model developed in the Voth group \cite{srivastavaHybridApproachHighly2013} simplify the lipid molecule description. While these allow for significantly larger spatial and temporal scales than atomistic modeling, they still explicitly simulate individual lipids. For example, the Martini model can effectively capture microsecond dynamics for assemblies of approximately $10^5-10^6$ lipids, \textit{i.e.}, roughly a square micrometer~\cite{pedersenMartini3Lipidome2025}. To access larger spatio-temporal length scales, necessary to accurately simulate the dynamics of entire organelle or Giant Unilamellar Vesicles (GUV) systems, we must shift to highly coarse-grained, or `mesoscopic', descriptions \cite{aytonNewInsightsBAR2009}. 

These mesoscopic models generally fall into two categories: mesh-based and particle-based. In both approaches, the fundamental discretization unit (a vertex or a bead) corresponds to a lipid patch several nanometers in size. Mesh-based models represent the membrane as a discretized, triangulated surface. These discretized models often employ a Monte Carlo scheme to sample the equilibrium configuration of the system~\cite{gompperMembranesFluctuatingTopology1998, dadunashviliFlippyUserFriendly2023, pezeshkianMesoscaleSimulationBiomembranes2024}. Conversely, solvent-free particle-based models at the mesoscopic level are based on molecular or Brownian dynamics~\cite{drouffeComputerSimulationsSelfAssembled1991, noguchiMeshlessMembraneModel2006,damaTheoryUltraCoarseGraining12013}. 
These models have been used to simulate large-scale planar membranes or vesicles \cite{yuanDynamicShapeTransformations2010,vanhille-camposModellingDynamicsVesicle2021}, interacting with colloidal particles~\cite{shenMembraneWrappingEfficiency2019} and coarse-grained filaments~\cite{harker-kirschneckChangesESCRTIIIFilament2019} or proteins~\cite{noguchiShapeDeformationLipid2016}. Despite sharing a common mesoscopic scale, the constitutive units of the two most used models differ fundamentally. The model of Yuan \textit{et al.}~\cite{yuanOneparticlethickSolventfreeCoarsegrained2010} utilizes a one-particle-thick representation, where a single anisotropic bead corresponds to a bilayer patch; its dynamics are governed by pairwise interactions scaled by a multiplicative orientation term. In contrast, Noguchi~\cite{noguchiSolventfreeCoarsegrainedLipid2011} models the bilayer as two coupled monolayers, employing a more complex additive potential that sums separate energetic terms for particle density and orientational alignment.
In this work, we introduce a new coarse-grained particle model operating in the mesoscopic regime, combining the simplicity of a single-particle-thick representation with an additive treatment of orientation-dependent interactions, thereby offering an alternative to the multiplicative scheme of Yuan \textit{et al.} and the bilayer-based construction of Noguchi. Our motivation arises from the need for a computationally efficient single-bead description of the membrane bilayer, the intention to provide an openly available and easily adoptable implementation, and the objective of establishing an unbiased potential form for particle interactions.

In this paper we present the formulation and derivation of the model and provide an implementation for the popular Molecular Dynamics (MD) engine LAMMPS~\cite{thompsonLAMMPSFlexibleSimulation2022}. We will demonstrate our model's capability to capture essential membrane behaviors, including self-assembly and the correct prediction of the fluctuation spectrum for planar membranes.

\section{Methods} \label{sec:methods}

\subsection{Model potential}
\begin{figure}[ht]
    \includegraphics[width=0.45\textwidth]{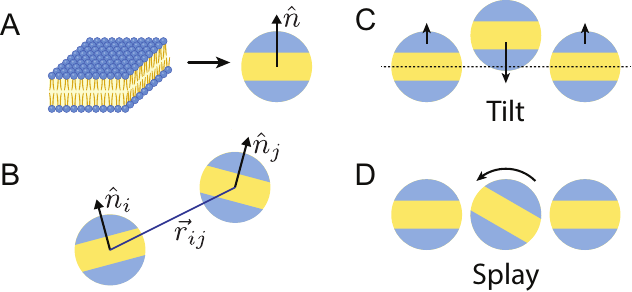}
    \centering
    \caption{Schematic of the model and interaction geometry. (A) Representation of a lipid bilayer patch as a single coarse-grained particle. The particle has rotational symmetry. (B) The separation vector $\vec{r}_{ij}$, and normal vectors $\hat{n}_i$ and $\hat{n}_j$ used in the derivation of the potential. (C) Schematic illustrating the forces resulting from a tilt deformation. (D) Schematic illustrating the torques resulting from a splay deformation.}
    \label{fig:schematic}
\end{figure}

The membrane is represented as a one-particle-thick sheet of coarse-grained elements. Each particle corresponds to a  patch of lipid bilayer with an area of a few nanometer squared and is modeled as a point-like dipole with a orientation vector $\hat{n}_i$ representing the local membrane normal. The model is additive in form and the solvent is modeled implicitly. All quantities are expressed in Lennard Jones (LJ) reduced units, with particle mass $m=1$, particle diameter $\sigma$ and energy scale $\varepsilon$ (the depth of the LJ well). Tilt $k_{\text{tilt}}$ and splay  $k_{\text{splay}}$ moduli are expressed in units of energy $\varepsilon$. The time unit will be $\tau_{\text{LJ}} = \sqrt{\varepsilon / m \sigma^2}$.

The total energy is written as

\begin{align}
\label{modeltotalenergy}
H &= U_{\text{rep}} + U_{\text{attr}} + U_{\text{tilt}} + U_{\text{splay}} \nonumber \\
  & = \sum_{i<j} \Big\lbrace U_{\text{rep}}(r_{ij}) + U_{\text{attr}}(r_{ij})  \nonumber \\
  &\quad + \Big[U_{\text{tilt}}(\hat{n}_i,\hat{n}_j,\hat{r}_{ij}) + U_{\text{splay}}(\hat{n}_i,\hat{n}_j) \Big] w(r_{ij}) \Big\rbrace,
\end{align}
where $\vec{r}_{ij} = \vec{r}_i - \vec{r}_j$, $r_{ij} = |\vec{r}_{ij}|$, and $\hat{r}_{ij} = \vec{r}_{ij}/r_{ij}$.

\begin{figure}[ht]
    \centering
    \includegraphics[width=0.45\textwidth]{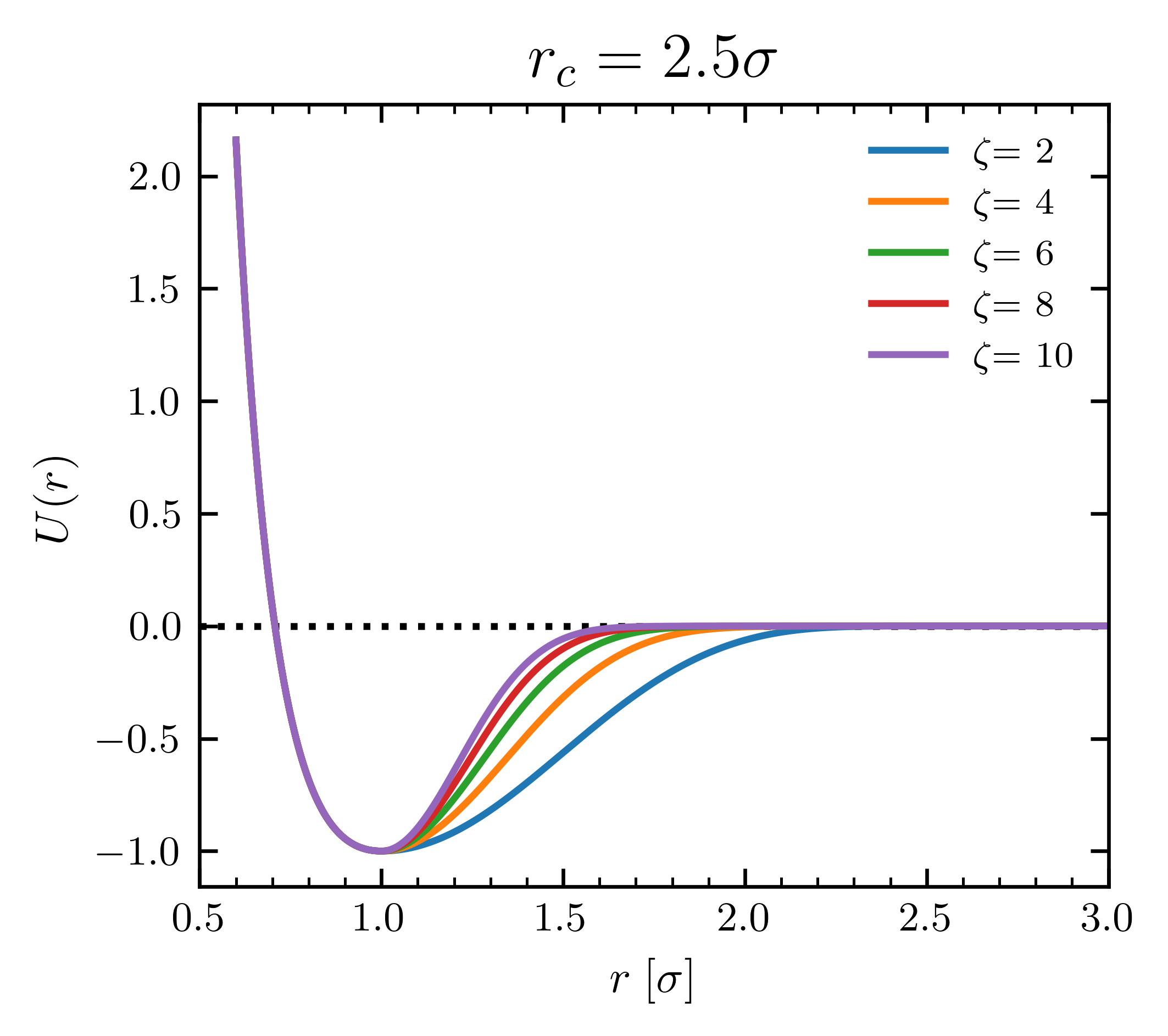}
    \caption{Isotropic potential $U(r)$ with isotropic cutoff $r_{\text{c}}=2.5$. The $\zeta$ parameter is tuning the slope of the attractive branch of the potential.}
    \label{fig:iso_potential}
\end{figure}

\medskip
\paragraph{Isotropic interactions.}  
Particles interact through a soft core excluded-volume repulsion combined with a short range attraction, see Fig.~\ref{fig:iso_potential}. The repulsive branch is given by a 4--2 LJ potential,
\begin{equation}
\label{Urep}    
U_{\text{rep}}(r) = \varepsilon \left[ \left( \frac{r_{\min}}{r} \right)^{4} - 2 \left( \frac{r_{\min}}{r} \right)^{2} \right], 
\qquad r \leq r_{\min},
\end{equation}
with cutoff at the potential minimum $r_{\min} = 1.0\;\sigma$. The softer 4--2 form, compared to the conventional 12--6 LJ potential, avoids crystallization behavior.

The attractive branch is described by a smoothly truncated cosine-squared potential,
\begin{equation}
\label{Uattr}
U_{\text{attr}}(r) =
\begin{cases}
- \varepsilon \cos^{2 \zeta} \left( \dfrac{\pi (r - r_{\min})}{2 \,(r_c - r_{\min})} \right), & r_{\min} < r < r_{\text{c}}, \\[1ex]
0, & r \geq r_{\text{c}},
\end{cases}
\end{equation}
which guarantees a continuous decay to zero at $r_{\text{c}}$ and avoids force discontinuities. The exponent $\zeta$ tunes the slope and the width of the attractive branch of the potential. The cut-off distance is also a model parameter tuning system behavior; in standard conditions we set its value to $r_{\text{c}}=2.5 \sigma$, such that each particle is interacting with two shells of neighbors.
Together these terms provide tunable short-range attraction with soft excluded volume, consistent with the previous one-particle-thick membrane model of Yuan \textit{et al.}~\cite{yuanOneparticlethickSolventfreeCoarsegrained2010}.

\medskip
\paragraph{Anisotropic interactions.} We introduce local orientational elasticity through tilt and splay interactions, which serve as a discretized version of the tilt model~\cite{hammElasticEnergyTilt2000}. The tilt term penalizes the deviation of particle directors from the local surface normal, whereas the splay term suppresses misalignment between neighboring particles.

The tilt energy for a particle pair is defined as:
\begin{equation}
\label{Utilt}
U_{\text{tilt}} = \frac{k_{\text{tilt}}}{2} \left[(\hat{n}_i \cdot \hat{r}_{ij})^2 + (\hat{n}_j \cdot \hat{r}_{ij})^2 \right].
\end{equation}
This potential is minimized when particle orientations are orthogonal to the interparticle axis $\vec{r}_{ij}$, consistent with a flat bilayer configuration. Physically, this interaction generates both translational forces and rotational torques.

To ensure orientational coherence across the bilayer, the splay energy is given by:
\begin{equation}
\label{Usplay}
U_{\text{splay}} = \frac{k_{\text{splay}}}{2}\left(\hat{n}_i \cdot \hat{n}_j -1 \right)^2.
\end{equation}
This term favors parallel (or antiparallel) alignment of neighboring dipoles. Unlike the tilt term, the splay interaction generates only torques.

\medskip
\paragraph{Weighting function.}  
Tilt and splay interactions act only within a finite range and are modulated by a smooth cutoff function \cite{noguchiSolventfreeCoarsegrainedLipid2011}:
\begin{equation}
\label{tiltsplaycutoff}
w(r) =
\begin{cases}
\exp \!\left( \dfrac{(r/r_{\text{ga}})^2}{(r/w_{\text{c}})^n - 1} \right), & r \leq w_{\text{c}}, \\[1ex]
0, & r \geq w_{\text{c}},
\end{cases}
\end{equation}
typically with $n=4$ and $r_{\text{ga}}=1.5\sigma$. The freely adjustable parameter $w_{\text{c}}$ is used to set the effective range of orientational interactions.

\medskip
\paragraph{Implementation.}  
Forces and torques are evaluated pairwise, with explicit derivations provided in the supplementary material. The model has been implemented in \textsc{LAMMPS}~\cite{thompsonLAMMPSFlexibleSimulation2022} writing a novel custom pair-style. The source code is openly available in our Gitlab repository: \href{https://gitlab.tudelft.nl/idema-group/MesoMem}{https://gitlab.tudelft.nl/idema-group/mesomem}.

\subsection{Langevin dynamics}
We simulated the membrane in the NVT ensemble (constant number of molecules N, volume V, and temperature T) with periodic boundary conditions in a box with sides of length $L_x$, $L_y$ and $L_z$.

To model the effect of an implicit solvent, we employed Langevin dynamics, in which translational and rotational motion are damped by frictional forces and subject to stochastic thermal noise. The damping serves two purposes: (i) it mimics the viscous drag exerted by the solvent on the particles, and (ii) it stabilizes particle orientations, ensuring correct alignment and structural integrity of the membrane.

The translational dynamics of each particle is described by
\begin{equation}
m \ddot{\vec{r}}_i = - \vec{F}_i - \gamma_t \dot{\vec{r}}_i + \vec{\eta}^{\;\text{t}}_i ,
\end{equation}
where $m$ is the particle mass, $\vec{F}_i$ are conservative forces, $\gamma_t$ the translational friction coefficient, and $\vec{\eta}^{\;\text{t}}_i$ a Gaussian random force with zero mean and variance determined by the fluctuation-dissipation relation.

The rotational dynamics uses a rotation vector representation (as implemented in \textsc{LAMMPS}) to integrate the unit orientation $\hat{n}_i$:
\begin{equation}
I \ddot{\hat{n}}_i = - \vec{T}_i - \gamma_r \dot{\hat{n}}_i + \vec{\eta}^{\;\text{r}}_i ,
\end{equation}
with $I = \frac{2}{5} m \sigma^2$ the moment of inertia of a spherical particle of radius $\sigma$, $\vec{T}_i$ are conservative torques, $\gamma_r$ the rotational friction coefficient, and $\vec{\eta}^{\;\text{r}}_i$ a Gaussian random torque. The rotation vector formalism ensures that the orientation remains normalized during integration.

We model the thermal bath using Gaussian white noise. Both the random forces and torques have a zero mean, meaning they do not drive the system in any specific direction: 
\begin{align}
    \langle \eta^{\text{t}}(t) \rangle &= 0, \nonumber \\ 
    \langle \eta^{\text{r}}(t) \rangle &= 0.
\end{align}

The strength of these fluctuations balances the dissipation, satisfying the standard fluctuation-dissipation theorem. For the translational noise $\eta^{\text{t}}$ and rotational noise $\eta^{\text{r}}$, the correlations are:
\begin{align}
\langle \eta^{\text{t}}(t) \, \eta^{\text{t}}(t') \rangle &= 2 k_\mathrm{B} T \, \gamma^{\text{t}} \, \delta(t-t'), \nonumber \\
\langle \eta^{\text{r}}(t) \, \eta^{\text{r}}(t') \rangle &= 2 k_\mathrm{B} T \, \gamma^{\text{r}} \, \delta(t-t').
\end{align}
Here, $\gamma^{\text{t}}$ and $\gamma^{\text{r}}$ represent the translational and rotational damping coefficients, respectively.

\subsection{Spontaneous curvature \texorpdfstring{$C_0$}{}}
\label{sec:spont_curv}
To account for the behavior of lipids with intrinsic curvature or leaflet asymmetry, we introduce a non-zero spontaneous curvature $C_0$. In our model, this is implemented by biasing the tilt and splay interactions to favor a preferred angle between neighboring particles, rather than the parallel alignment we obtain in the absence of spontaneous curvature. For a given value of $C_0$, the preferred angle between particles depends on their distance $r_{ij}$ according to $\sin(\frac{\alpha}{2}) = \frac{1}{2} r_{ij} C_0$ (see Fig.~\ref{fig:S2} and supplementary Sec.~\ref{sec:curvature}).
The interaction potentials are modified to penalize deviations from this preferred orientation. The updated energy terms for tilt and splay are given by:

\begin{align}
U_{\text{tilt}} &= \frac{k_{\text{tilt}}}{2} \left[ \left(\hat{n}_i \cdot \hat{r} + \frac{1}{2} r C_0 \right)^2 + \left(\hat{n}_j \cdot \hat{r} - \frac{1}{2} r C_0\right)^2 \right], \\
U_{\text{splay}} &= \frac{k_{\text{splay}}}{2} \left[\hat{n}_i \cdot\hat{n}_j - 1 + 2\, \left(\frac{1}{2}C_0 r\right)^2 \right]^2.
\end{align}

\subsection{Implementation in LAMMPS} We implemented Langevin dynamics using a combination of the \textsc{nve} and \textsc{langevin} fixes in LAMMPS. Together, these commands function as a modified Verlet integrator, incorporating both the random and frictional forces necessary to solve the Langevin equations of motion.

Unless stated otherwise, our simulation parameters are:

\begin{itemize}
    \item LJ units: $\sigma=1.0$, $\varepsilon=1.0$,
    \item timestep: $\Delta t = 0.01 \; \tau_{\text{LJ}}$,
    \item translational damping coefficient: $\gamma_t = 1.0 \; \tau_{\text{LJ}}^{-1}$,
    \item rotational damping coefficient: $\gamma_r = 1.6 \; \sigma^2\tau_{\text{LJ}}^{-1}$.
\end{itemize}

We performed benchmarks to test parallelization and scalability of our systems (see Fig.~\ref{fig:S8} in the supplementary material).

Examples of LAMMPS scripts are publicly available in our Gitlab repository.

\begin{figure*}
    \includegraphics[width=\textwidth]{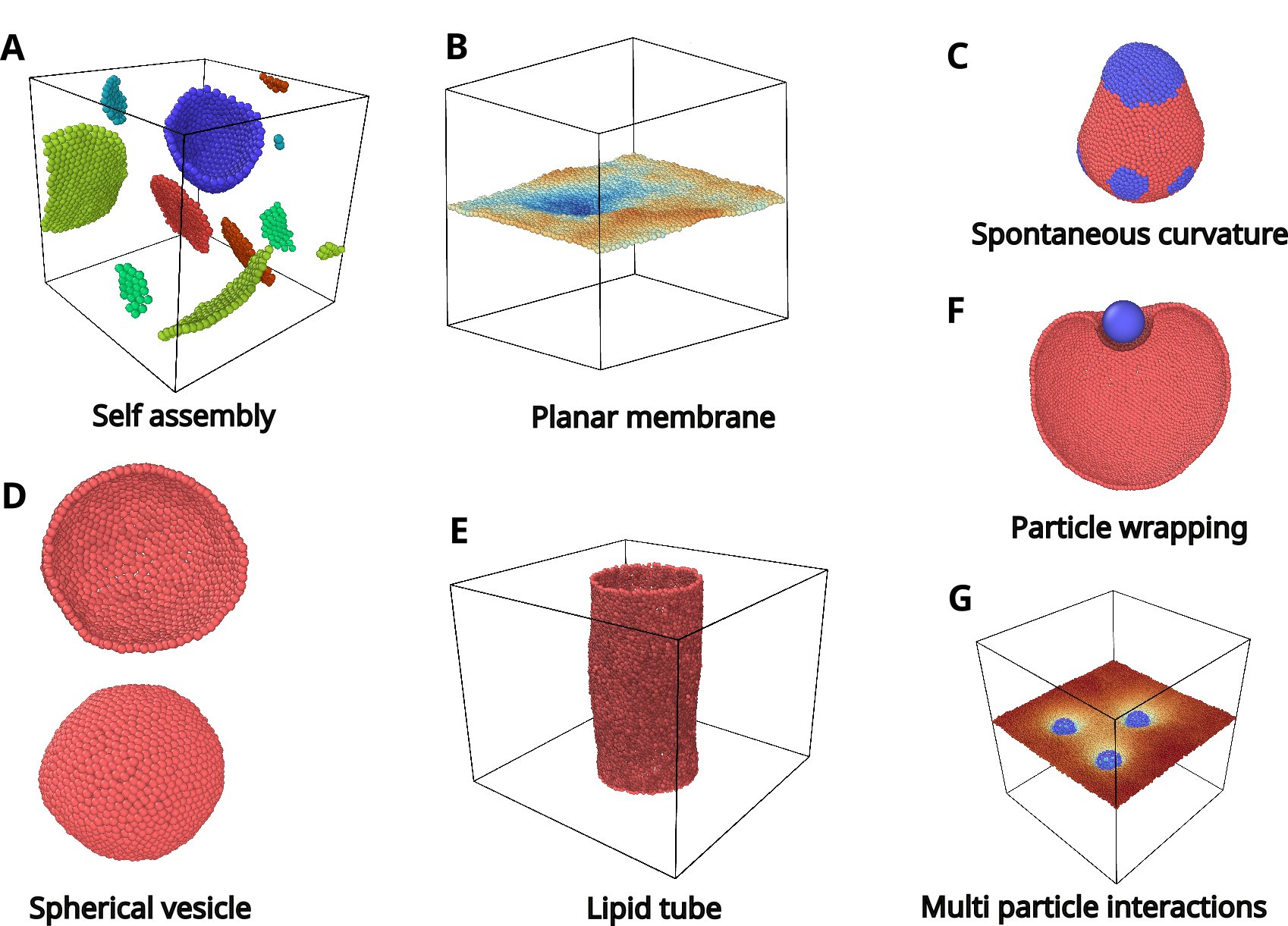}
    \centering
    \caption{Overview of simulated lipid systems. (A) Self-assembled patches formed from 1500 randomly placed particles. (B) Planar membrane colored according to $z$-height. (C) Lipid vesicle composed of zero spontaneous curvature beads (in red) and non-zero spontaneous curvature $C_0 = 0.1 \sigma^{-1}$ beads (in blue). (D) Spherical vesicle shown as an $xz$-cross-section (top) and whole vesicle (bottom).  (E) Membrane tube. (F) Cross-section ($xz$-plane) of a spherical vesicle wrapping a particle. (G) Planar membrane interacting with multiple soft colloidal metaparticles~\cite{paesaniMetaparticlesComputationallyEngineered2024}. Initial configurations have been generated following the procedures described in Sec.~\ref{sec:init}.}
    \label{fig:panel_systems}
\end{figure*}

\section{Results \& Discussion} \label{sec:Discussions}
We begin by establishing the feasibility of self-assembly and the conditions under which a membrane is stable, ensuring the system reaches a robust equilibrium configuration capable of maintaining the layer's integrity. Following upon the structural integrity, we characterize the membrane's phase behavior and lipid diffusivity. We use the results of the diffusivity measurement to map simulation results to physical units, which allows us to directly measure physical quantities.
Considering the membrane as a material, we measure its constitutive properties, specifically the area compressibility and bending modulus, which dictate the membrane's energetic response to mechanical stress and thermal fluctuations.

Finally, as example applications, we study the wrapping of a solid particle by the membrane and we simulate the membrane reshaping after an osmotic shock. Fig.~\ref{fig:panel_systems} shows an overview of our simulation results.

\subsection{Self-assembly}
As the first step in the validation of the model, we determined whether our simulated particles can spontaneously assemble into lamellar structures from a disordered state. We initialized the system with a random configuration of $N=1500$ particles in a cubic box of side length $L=20\sigma$. This configuration corresponds to a reduced volume fraction of $\phi = N V_p / V_{\text{box}} \approx 0.1$, where $V_p = \frac16 \pi \sigma^3$ is the volume of a single particle. Particles started forming small patches at intermediate stages ($t=500\;\tau_{\text{LJ}}$), as can be seen in Fig.~\ref{fig:S3}. The membrane patches then coalesce into large planar membranes around  $t=2000\;\tau_{\text{LJ}}$.

Fig.~\ref{fig:S4} displays representative snapshots of the system at an intermediate stage ($t=500\;\tau_{\text{LJ}}$) across a range of tilt moduli $k_{\text{tilt}}$. Secondary model parameters were held fixed at $k_{\text{splay}}=1$, $\zeta=5.0$, $w_{\text{c}}=2.0 \sigma$ and a cutoff radius $r_{\text{c}}=2.5 \sigma$. We find that $k_{\text{tilt}}$ is the critical control parameter governing the morphology of the assembled structures. At values below $k_{\text{tilt}} \approx 10.0$, the energetic penalty for director misalignment is too weak to maintain planar order, leading instead to compact, isotropic aggregates. Once $k_{\text{tilt}}$ surpasses this threshold, increased orientational stiffness drives these clusters to flatten into membrane-like patches. Unless otherwise noted, $k_{\text{tilt}}$ is set to a standard value of 12.0. These patches continue to grow and coalesce into large planar membranes at $t=2000\;\tau_{\text{LJ}}$.

The morphology of growing membrane patches is governed by the competition between the bending energy, which favors a flat geometry (for zero spontaneous curvature $C_0=0$), and the line tension, which penalizes the open edges of the patch. Vesiculation occurs when the energetic gain from eliminating the edge exceeds the cost of bending the membrane into a closed shell. To observe this transition, we extended the simulation time to $3500\,\tau_{\text{LJ}}$, at which point we observed the spontaneous formation of closed vesicles, see Fig.~\ref{fig:S3}. To distinguish between kinetic assembly and thermodynamic stability, we compared these results to the behavior of pre-formed structures. We find that the larger spherical shells described in Sec.~\ref{sec:init} (Fig.~\ref{fig:panel_systems}D) remain structurally intact and stable under identical simulation conditions. This observation confirms that the model is capable of maintaining stable vesicular morphologies.

\subsection{Conditions of stability}
\label{sec:stabilityconditions}
To test whether our interaction scheme supports a stable planar membrane, we prepared initial configurations by placing particles on a hexagonal lattice with spacing $a = 0.8\sigma$. A lattice of $50 \times 50$ sites was used as benchmark size: it is large enough to limit finite-size effects while remaining computationally feasible. Simulations were performed using Langevin dynamics as detailed in Sec.~\ref{sec:methods}. We introduced a barostat as well, to ensure that the lateral pressure in the membrane vanished, \textit{i.e.}, $P_{xx} = P_{yy} = 0$, thereby ensuring a tension-free equilibrium state. Each system was simulated for $5\times 10^{4}$ steps with time step $\Delta t = 0.01 \tau_{\text{LJ}}$, which allows the planar membrane to relax from the initial condition. 
Stability is evaluated by monitoring particle clustering. A simulation is considered stable if, at the end of the run, the majority (more than 99\%) of particles belong to a single bigger cluster corresponding to the planar membrane. We tolerated single particle evaporation events, as long as the planar membrane remained intact and continuous. This criterion was checked using a clustering algorithm, the results are shown in Fig.~\ref{fig:S5}.

Our parameter sweep yielded the following observations:

\begin{itemize}
    \item \textbf{Interaction cutoff ($r_{\text{c}}$):} A minimum cutoff of $r_{\text{c}} = 2.5 \sigma$ is required to sustain aggregation. Beyond this value, the membrane stability is largely insensitive to further increases in interaction range.
    \item \textbf{Orientation cutoff ($w_{\text{c}}$):} This parameter is effectively upper-bounded by $r_{\text{c}}$. Variations below this bound showed negligible impact on structural stability (see Fig.~\ref{fig:S5}) but a significant influence on the mechanical stiffness of the membrane as discussed in Sec.~\ref{sec:fluctuationspectrum} below.
    \item \textbf{Tilt modulus ($k_{\text{tilt}}$):} A value of $k_{\text{tilt}} > 10.0 $ is strictly necessary to stabilize the planar topology against collapse into isotropic clusters. Above this critical stiffness, the membrane remains robust.
    \item \textbf{Splay modulus ($k_{\text{splay}}$):} Interestingly, the splay constant $k_{\text{splay}}$ exhibits significant influence on the stability of the flat membrane. Most of the simulations have been run with a range  $0.5 \le k_{\text{splay}} \le 1.0 $. Values above this range disrupt the stability of the membrane. 
    \item \textbf{Attraction slope ($\zeta$):} For low values $\zeta \ll 6.0$ the systems is jammed and solid-like due to the longer range of attraction (see Sec.~\ref{sec:phasediagram}).
    \item \textbf{Temperature ($T$):} Although not an explicit parameter of the force field, the temperature together with $\zeta$ acts as the principal driver of phase behavior. At low $T$ we obtain a crystalline or gel-like planar membrane, while increasing $T$ induces a fluid phase, eventually leading to complete sublimation (evaporation) of the bilayer at high $T$.
\end{itemize}
The phase diagram (Fig.~\ref{fig:phase_plot}) highlights the effects of $\zeta$ and $T$ values leading to distinct gel, liquid, and gas regions.

\subsection{Spontaneous curvature}

Fig.~\ref{fig:panel_systems}C depicts a vesicle composed of two bead types: type A (in red) with zero spontaneous curvature ($C_0=0$) and type B (in blue) with positive spontaneous curvature ($C_0=0.1 \sigma^{-1}$). A vesicle consisting of an initially random mixture of type A and type B particles undergoes phase separation and changes it shape to minimize the bending and mixing energy. While not the primary focus of this study, this example demonstrates how spontaneous curvature and different lipid types can be incorporated easily into our framework.

\subsection{Membrane characterization}
Following the stability assessment of the membrane layer, we characterized its fundamental physical properties and their dependence on system parameters. We focused on key metrics standard in computational membrane models: the phase diagram, line tension, lipid lateral diffusivity, area compressibility, and bending modulus. We evaluated these properties using the previously described system, consisting of a planar square membrane of 2500 particles under periodic boundary conditions and run for $10^5$ steps.

\subsubsection{Phase diagram and lipid diffusivity}
\label{sec:phasediagram}
\begin{figure}[ht]
    \centering
    \includegraphics[width=\columnwidth]{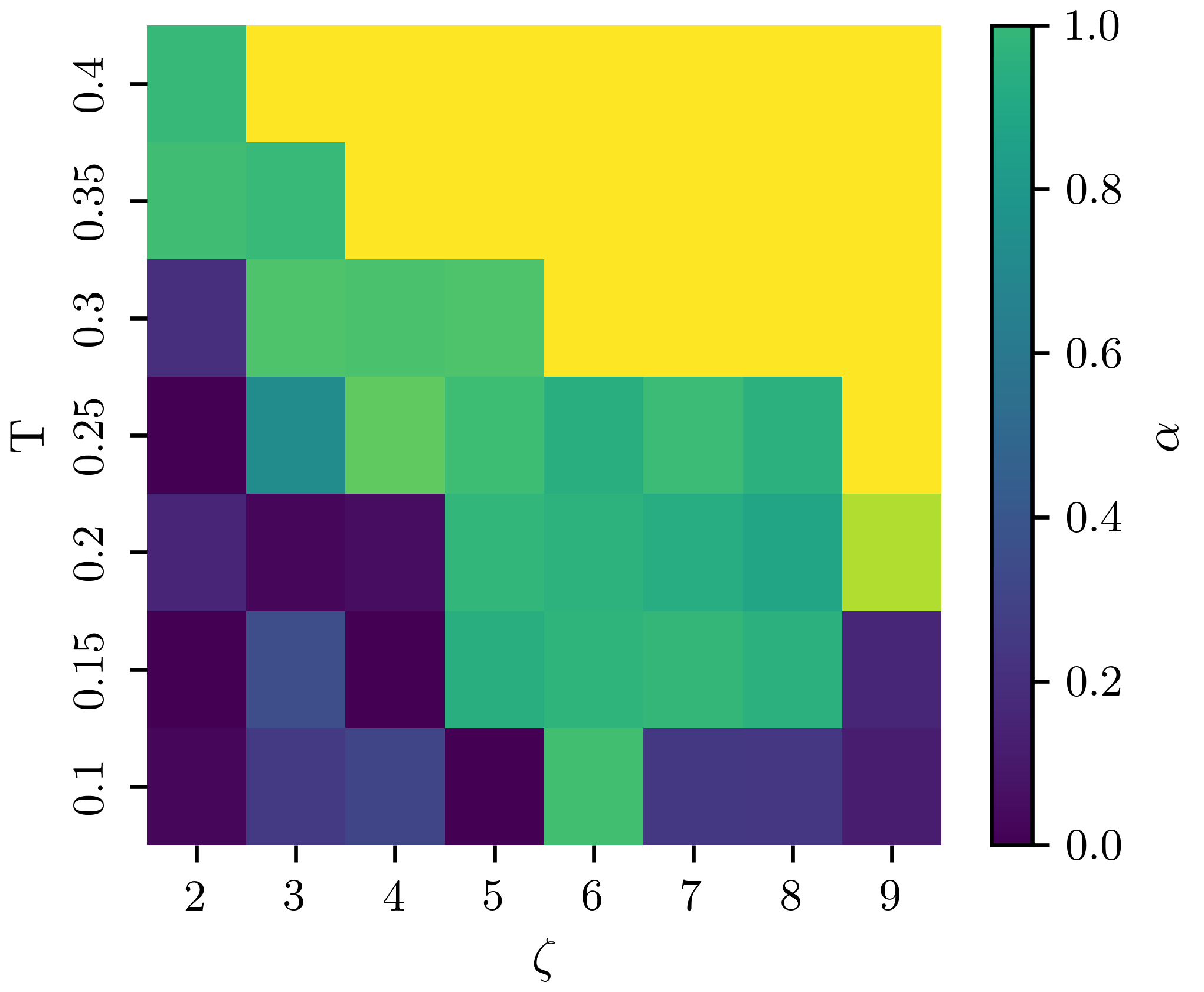}
    \caption{Dynamical phase diagram of the membrane model in the ($T$, $\zeta$) parameter space. The color map represents the scaling exponent $\alpha$ obtained from the power-law fit of the lipid Mean Squared Displacement ($\text{MSD} \propto t^\alpha$). We indentify three distinct regimes: a sub-diffusive solid/gel phase ($\alpha \ll 1$), a diffusive fluid phase ($\alpha \approx 1$), and a gas-like regime ($\alpha \gg 1$) corresponding to membrane disintegration. Fixed model parameters: $N=2500$, $k_{\text{tilt}}=12$, $k_{\text{splay}}=1$, $w_{\text{c}}=2.0\sigma$ and $r_{\text{c}}=2.5 \sigma$.}
    \label{fig:phase_plot}
\end{figure}

To characterize the physical state of the membrane, we analyzed the scaling behavior of the lipid Mean Squared Displacement (MSD), characterized as $\text{MSD}(t) \propto t^\alpha$. By plotting the exponent $\alpha$ as a heat map in the ($T$, $\zeta$) parameter space (Fig.~\ref{fig:phase_plot}), we can easily identify three distinct dynamical regimes. The system exhibits a clear separation between a sub-diffusive solid/gel phase ($\alpha \ll 1$), a diffusive liquid fluid phase ($\alpha \approx 1$), and a gas-like regime (in yellow) corresponding to membrane disintegration.

The phase behavior is primarily governed by the temperature $T$ and the interaction parameter $\zeta$, which controls the attractive slope of the inter-particle potential. An increase of $\zeta$ at higher temperature $T$ leads to a transition to the gas phase. This counter intuitive behaviour can be explained by the fact that a higher value of $\zeta$ not only gives rise to stronger restoring forces but also effectively reduces the attractive range as well. Secondary model parameters were held fixed at $k_{\text{tilt}}=12$, $k_{\text{splay}}=1$, $w_{\text{c}}=2.0 \sigma$ and a cutoff radius $r_{\text{c}}=2.5 \sigma$.

\begin{figure}[ht]
    \centering
    \includegraphics[width=\columnwidth]{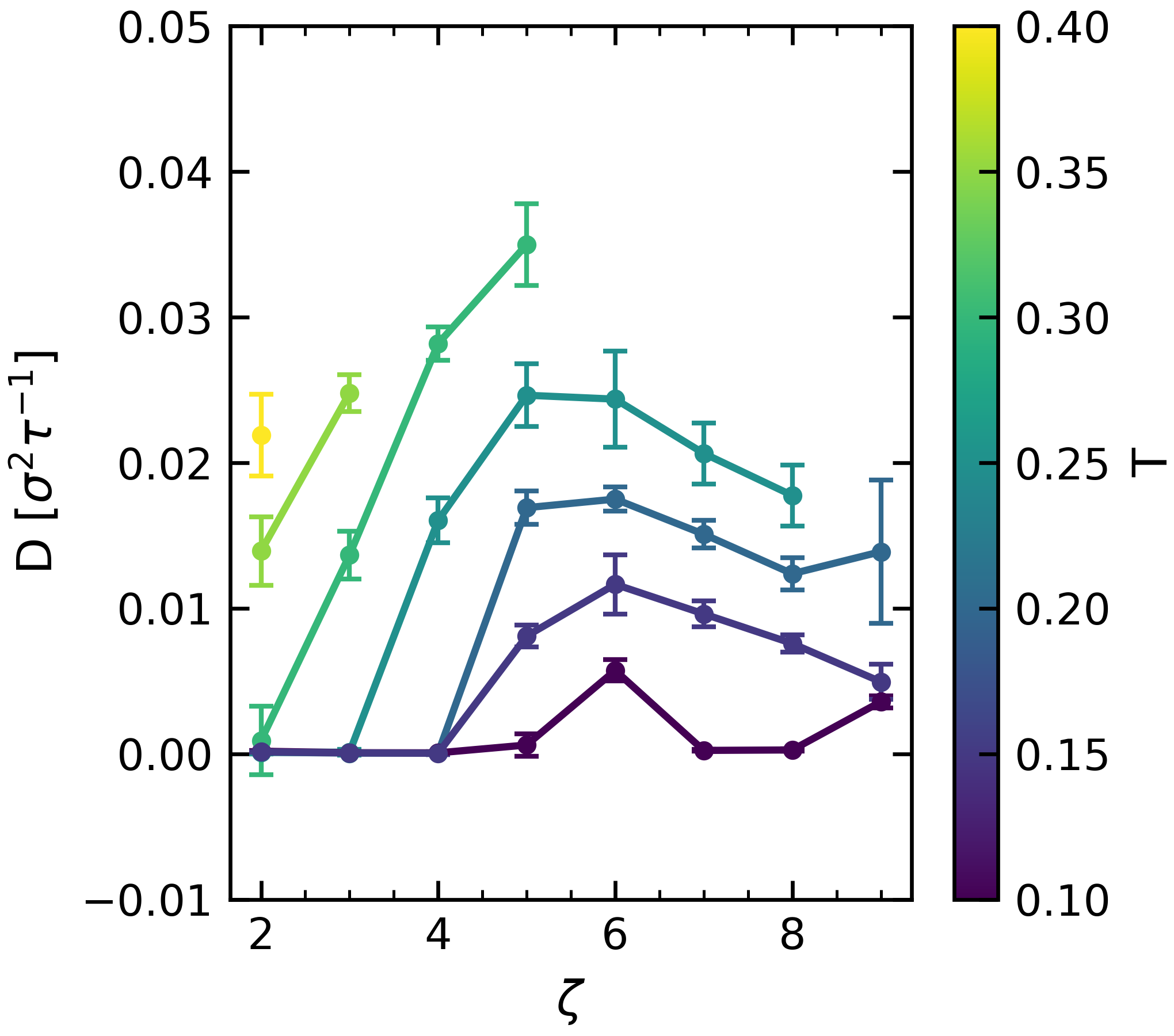}
    \caption{Lateral diffusion coefficient $D$ of membrane lipids as a function of temperature $T$ and interaction parameter $\zeta$, from the same simulations as used for Fig.~\ref{fig:phase_plot}. $D$ was calculated by fitting the MSD to $\text{MSD} = 4Dt$ in the linear regime. Points of the dataset corresponding to a system in the gas phase have been excluded from the plot. Errorbars are standard deviations over 3 independent replicas.}
    \label{fig:diffcoeff_zeta}
\end{figure}

Following the phase analysis, we quantified the lipid dynamics within the stable membrane regimes. We determined the lateral diffusion coefficient $D$ by fitting the linear regime of the MSD to the relation for two-dimensional diffusion, $\text{MSD} = 4Dt$. Data points corresponding to the gas phase were excluded from this analysis as the diffusive model is not applicable to free particles. Fig.~\ref{fig:diffcoeff_zeta} shows the dependence of lipid diffusivity on $T$ and $\zeta$.

\subsubsection{Real world unit mapping}
To relate the simulation parameters to physical scales, we map the model units $\sigma$ and $\tau_{\text{LJ}}$ to the properties of a typical phospholipid membrane.
Equating our particle size $\sigma$ to the lipid bilayer thickness ($5\;\text{nm}$) and given that a single lipid molecule occupies an area of $0.5 \; \text{nm}^2$, it follows that each particle in our membrane model represents $\approx 80$ lipids ($40$ in each leaflet).

Unlike its readily defined length scale, a membrane system lacks a uniquely define d time scale for its dynamics. Instead, the time scale is typically derived from the lipid diffusion constant. Given a characteristic experimentally measured lateral diffusion constant $D \approx 1\;\mu \text{m}^2 / \text{s}$~\cite{faheyLateralDiffusionPhospholipid1978} and a measured simulation diffusion constant $D_{\text{sim}} = 0.025 \sigma^2 / \tau_{\text{LJ}}$, the simulation time unit maps to $\tau_{\text{LJ}} \approx 0.625 \; \mu \text{s}$.

\subsubsection{Line tension} \label{sec:linetension}
A free edge of a membrane carries a line tension~$\lambda$. This interfacial penalty arises because particles situated at the membrane boundary experience a different local environment and fewer cohesive interactions compared to those within the bulk. The line tension drives the system to minimize its exposed edges. This thermodynamic driving force governs macroscopic processes such as spontaneous vesiculation (see Fig. \ref{fig:S3}) and the circular conformation of free membrane patches.

To quantify~$\lambda$, we employ the methodology described in Ref.~\cite{tolpekinaSimulationsStablePores2004}. Specifically, we initialize a membrane strip, equilibrate it at zero surface tension, and compute the lateral stress profile across the periodic boundaries.

We systematically investigate how the simulation temperature~$T$ and the fundamental model parameters, the orientation cutoff~$w_{\text{c}}$ and the parameter~$\zeta$, influence the line tension, see Fig.~\ref{fig:tension}. Because line tension primarily originates from cohesive particle interactions, we find that varying $w_{\text{c}}$ does not significantly impact $\lambda$ (indicated by the constant horizontal lines in Fig.~\ref{fig:tension}). Conversely, we observe a strong temperature dependence, as expected. The temperature effectively scales the relative strength of the isotropic attraction well; thus, at lower temperatures, the cohesive isotropic forces dominate, yielding a higher line tension. Furthermore, the parameter~$\zeta$, which specifically controls the slope of the attractive potential, directly modulates the energetic cost of boundary formation and correspondingly affects the measured line tension.

\subsubsection{Area compressibility modulus}

The area compressibility modulus, $K_{\text{A}}$, quantifies the membrane's resistance to isotropic expansion. This parameter is determined by analyzing the stress-strain response of the system under lateral tension. In our simulations, we subjected a planar membrane (2500 particles) to equibiaxial stretching ($\gamma_{xx} = \gamma_{yy}$) across the $xy$-plane. We monitored the resulting surface tension, $\Sigma$, as a function of the projected area until the onset of mechanical failure, defined by the formation of a membrane pore.

\begin{figure}[ht]
    \centering
    \includegraphics[width=0.45\textwidth]{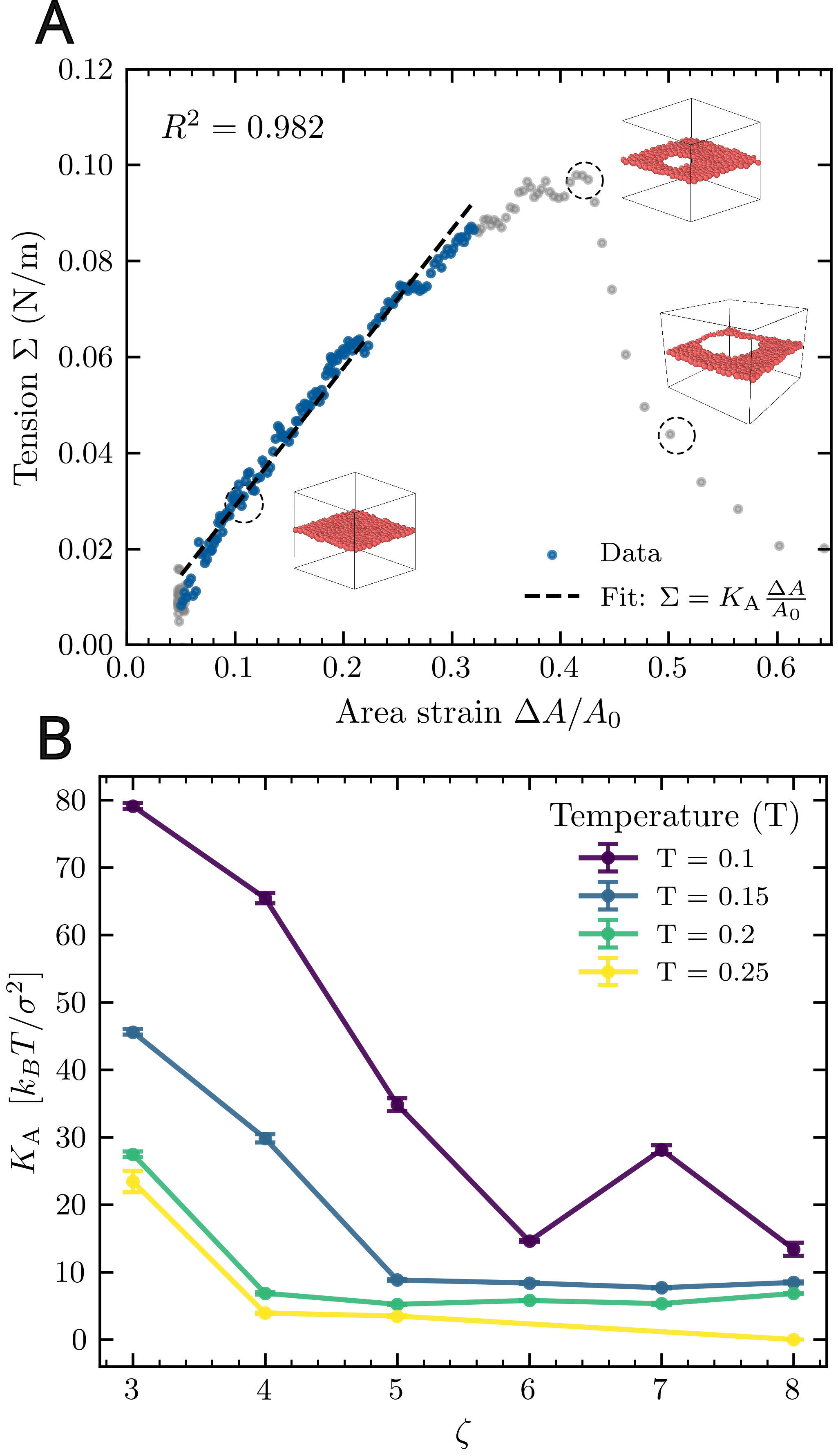}
    \caption{(A) Representative plot of tension $\Sigma$ versus area strain $(A - A_0)/ A_0$. The compressibility modulus $K_{\text{A}}$ is determined from the slope of the linear regime. Rupture occurs at a strain of $\approx 0.4$. The parameters for this simulation are $N=400, k_{\text{tilt}}=12, k_{\text{splay}}=1, \zeta=5.0, w_{\text{c}} = 2.0\sigma, T=0.3$. (B) Compressibility modulus $K_{\text{A}}$ as a function of temperature $T$ and slope parameter $\zeta$. The sharp decrease in $K_{\text{A}}$ corresponds to the transition from a solid-like to a fluid-like phase. The parameters for this dataset are $N=2500, k_{\text{tilt}}=12, k_{\text{splay}}=1, w_{\text{c}} = 2.0\sigma$}
    \label{fig:Compressibility}
\end{figure}

The modulus $K_{\text{A}}$ is derived from Hooke's law for elastic membranes in the low-strain regime:

\begin{equation}
    \Sigma = K_{\text{A}} \frac{A - A_0}{A_0}
    \label{eq:compressibility}
\end{equation}

where $A$ is the projected area under tension, $A_0$ is the equilibrium area at zero tension, and the term $(A - A_0)/A_0$ represents the area strain $\alpha$. By plotting the applied tension $\Sigma$ against the area strain, we can obtain $K_{\text{A}}$ from the slope of the linear fit (Fig.~\ref{fig:Compressibility}A).

The area compressibility modulus $K_{\text{A}}$ was extracted from the linear regime of the tension-strain curves prior to failure. Our analysis reveals distinct dependencies of $K_{\text{A}}$ on the thermodynamic phase of the system (gel, liquid, or gas). By comparing the compressibility data in Fig.~\ref{fig:Compressibility} with the phase diagram in Fig.~\ref{fig:phase_plot}, we observe a significant mechanical response at the phase boundaries. Specifically, a sharp decrease in $K_{\text{A}}$ occurs at the gel-to-liquid transition, signaling the loss of long-range translational order. As anticipated, the primary parameters influencing $K_{\text{A}}$ are those governing the phase behavior: the temperature $T$ and the attractive slope parameter $\zeta$. Both parameters strongly modulate $K_{\text{A}}$ by driving the phase transitions. High $K_{\text{A}}$ values are characteristic of the solid-like (gel) phase, whereas a marked reduction is observed in the fluid-like phase.


\subsubsection{Fluctuation spectrum and bending modulus}
\label{sec:fluctuationspectrum}
\begin{figure}[ht]
    \centering
    \includegraphics[width=\columnwidth]{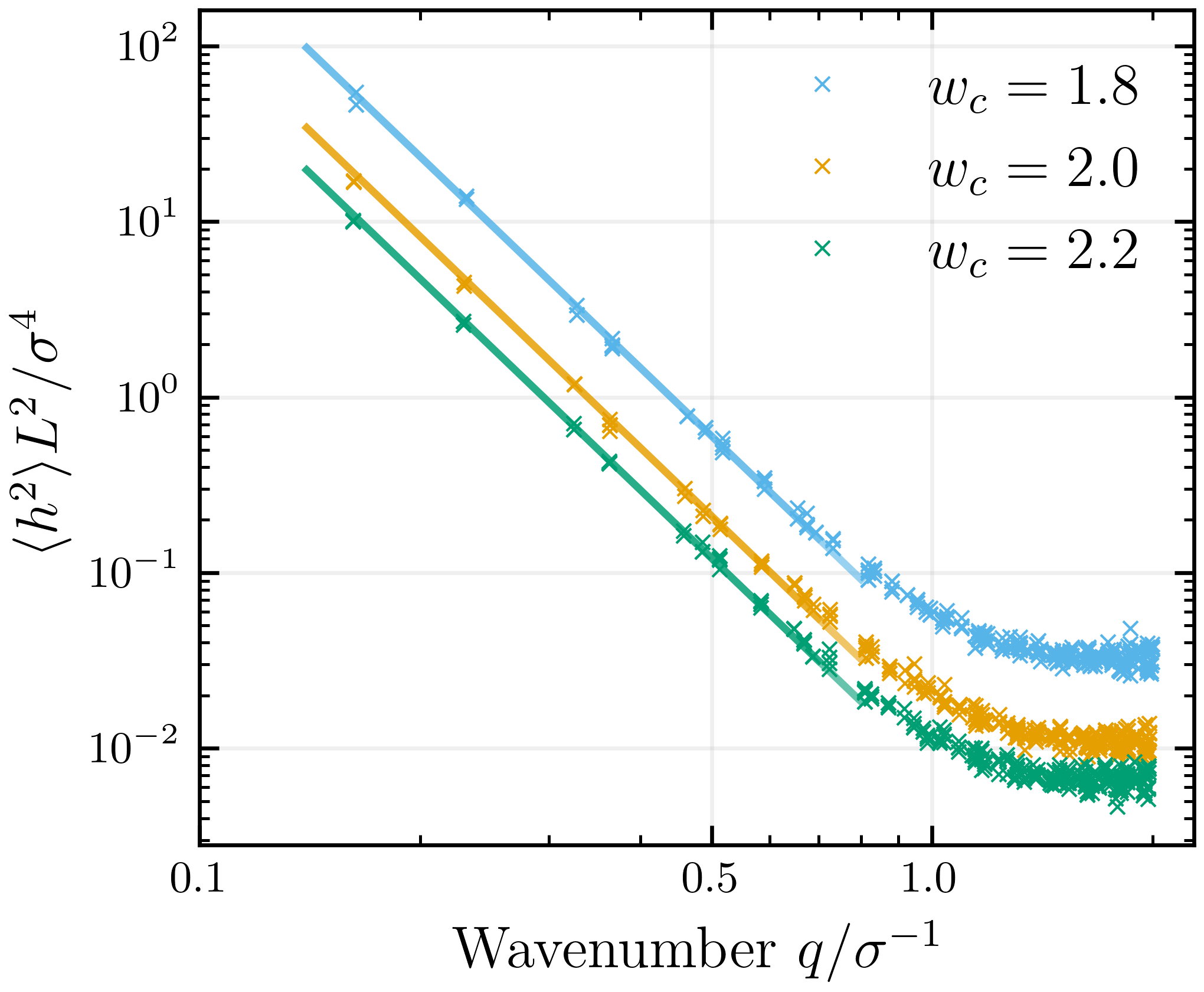}
    \caption{Power spectrum of membrane height fluctuations. The mean squared amplitude $\langle |h_q|^2 \rangle$ is plotted against the wave number $q$. The simulation data exhibits a clear $q^{-4}$ scaling behavior, consistent with the theoretical prediction for a tensionless membrane. This confirms that the system is in a zero-surface-tension state ($P_{xx} = P_{yy} = 0$), where fluctuations are governed solely by bending rigidity.}
    \label{fig:spectrum}
\end{figure}

\begin{figure}[ht]
    \centering
    \includegraphics[width=\columnwidth]{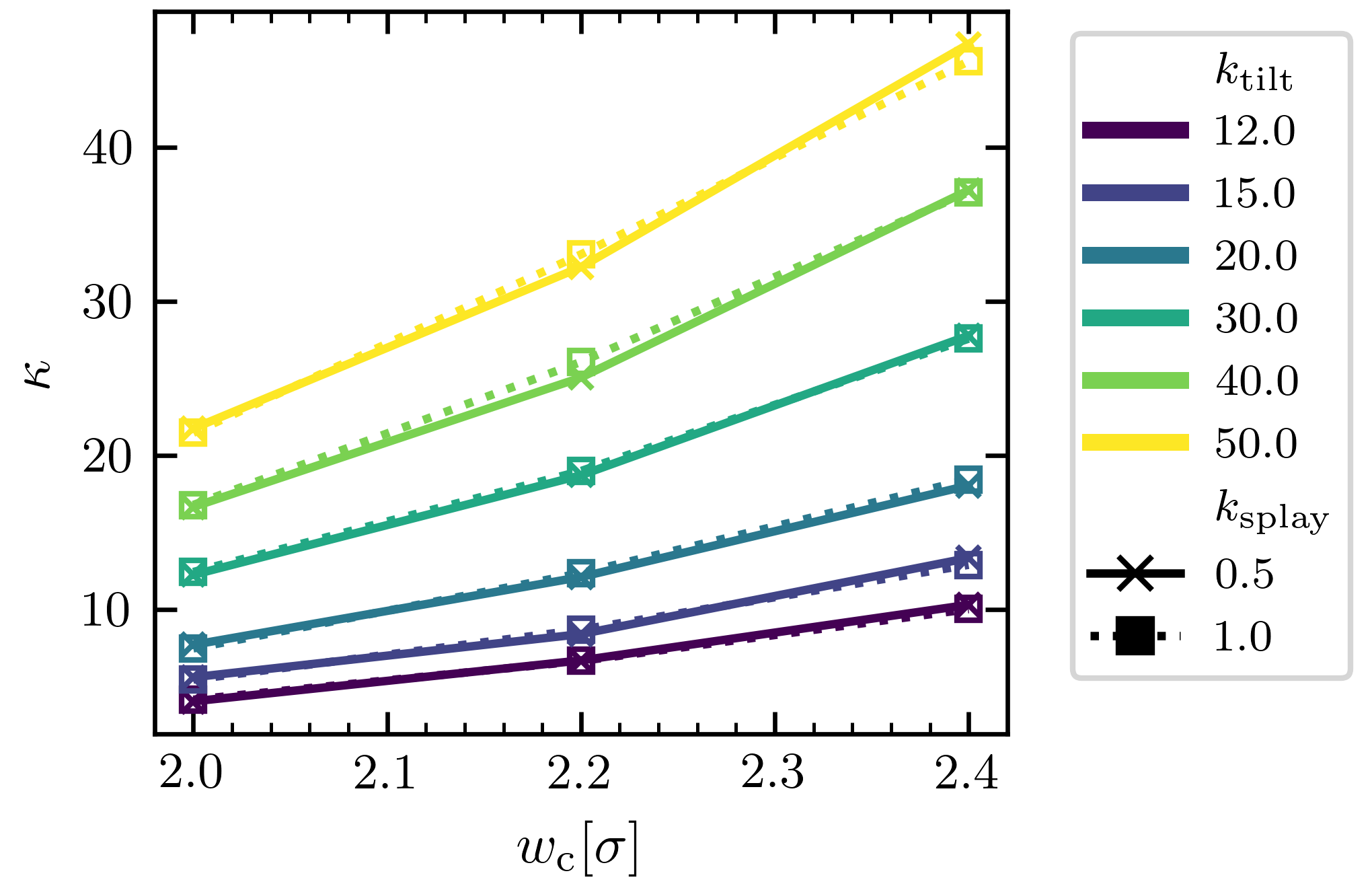}
    \caption{Dependence of the bending modulus $\kappa$ on model parameters. The membrane rigidity increases monotonically with the orientational cutoff $w_{\text{c}}$ and the tilt modulus $k_{\text{tilt}}$. In contrast, the splay modulus $k_{\text{splay}}$ has a negligible impact on the overall bending modulus.}
    \label{fig:kappa}
\end{figure}

According to continuum membrane theory, the height fluctuation spectrum in the tensionless regime is described by the Helfrich expression~\cite{helfrichUndulationsStericInteraction1984}:
\begin{equation}
    \langle |h_q|^2 \rangle = \frac{1}{L^2}\frac{k_\mathrm{B} T}{\kappa q^4},
    \label{eq:fluctuation_spectrum}
\end{equation}
where $k_\mathrm{B} T$ is the thermal energy, $L$ is the lateral box size, $\kappa$ is the bending modulus, and $q$ is the magnitude of the wave vector. To compute the spectrum and perform the fitting procedure, we adopted the analysis framework established by Muñoz-Basagoiti \textit{et al.}~\cite{munozbasagoitiTutorialMesoscaleComputer2025}, inspired by the error analysis protocols in~\cite{erguderIdentifyingSystematicErrors2021}. We adapted their implementation—originally designed for the Cooke-Deserno three-bead model—to accommodate the specific geometry of our membrane model. All reported values for $\kappa$ satisfy the goodness-of-fit criterion $Q > 10^{-4}$, as recommended in Ref.~\cite{erguderIdentifyingSystematicErrors2021}. The membrane was simulated following the stability protocols described in Sec.~\ref{sec:stabilityconditions}. System equilibration was verified by monitoring the convergence of the lateral box area ($L_x \times L_y$) to a steady-state plateau. To ensure statistically independent sampling, we calculated the time autocorrelation function for the Fourier modes. The decorrelation time for the lowest-$q$ mode was found to be approximately $1000 \, \tau_{\text{LJ}}$, which was subsequently used as the sampling interval. A total of 250 uncorrelated snapshots were used to compute the power spectrum. As illustrated in Fig.~\ref{fig:spectrum}, the fluctuation spectrum of the simulated membrane exhibits the characteristic $q^{-4}$ scaling expected for a tensionless interface. The bending modulus $\kappa$ was extracted by fitting the numerical data to Eq.~\ref{eq:fluctuation_spectrum}. As shown in Figs.~\ref{fig:spectrum} and \ref{fig:kappa}, the effective membrane stiffness is primarily governed by the orientational cutoff $w_{\text{c}}$ and the tilt modulus $k_{\text{tilt}}$, while the splay modulus $k_{\text{splay}}$ has only a minor effect. We confirmed that $\kappa$ is independent of both the coupling parameter $\zeta$ and the cutoff radius $r_{\text{cut}}$. Physically, increasing $w_{\text{c}}$ stiffens the membrane by incorporating a greater number of neighbors into the local orientation penalty. A similar stiffening effect is observed when increasing $k_{\text{tilt}}$. By tuning these parameters, the model accurately reproduces physiological bending moduli in the range of $10\text{--}30 k_\mathrm{B} T$~\cite{goetzMobilityElasticitySelfAssembled1999,dimovaRecentDevelopmentsField2014}.

\subsubsection{Equilibrium tube radius}
Under tension, membranes readily form long membrane tubes~\cite{Evans1996,derenyiFormationInteractionMembrane2002,powersFluidmembraneTethersMinimal2002,Koster2005}. When pre-arranged in tubular shape, our particles form a stable membrane tube under tension, see Fig.~\ref{fig:panel_systems}E. The relationship between the radius of a membrane tube and the applied lateral tension is a well-established result of continuum membrane theory~\cite{derenyiFormationInteractionMembrane2002}. It follows from minimizing the sum of the Helfrich bending energy and the membrane tension, which for a cylindrical shape is given by $E_\text{tube} = \pi\kappa L/R + 2\pi\gamma R L - fL$. Minimizing this energy with respect to the tube radius $R$ gives

\begin{equation}
    \label{equilibriumtuberadius}
    R_{\text{eq}} = \sqrt{\frac{\kappa}{2\gamma}}.
\end{equation}
The membrane tube is equilibrated under the action of a barostat, imposing different tension values. The equilibrium radius decreases with increasing membrane tension. The membrane tension can be obtained from measurement of the stress tensor calculated in LAMMPS as described in in the Supplementary Material and in \cite{munozbasagoitiTutorialMesoscaleComputer2025}.
We test this scaling across three values of the potential range
$w_{\text{c}} \in \{2.0,\,2.2,\,2.4\}$ and three tilt stiffnesses
$k_{\text{t}} \in \{15,\,20,\,30\}$ (see Sec.~\ref{sec:tuberadiusmeasurement} in the Supplementary Material for the simulation protocol). Across all parameter combinations the fitted exponent of the relation $R_{eq} \propto \gamma^{-\alpha}$ is consistent with the theoretical value $\alpha = 0.5$ given in equation~(\ref{equilibriumtuberadius}), confirming that the model correctly reproduces equilibrium tube mechanics, see Fig.~\ref{fig:tuberadius}.

\subsection{Particle wrapping}
\label{sec:wrapping}
To demonstrate the applicability of our model, we investigated the adhesive interaction between a colloidal particle and a fluid membrane, corresponding to frequently used adhesion assays in biophysical studies \cite{vanderwelLipidMembranemediatedAttraction2016} and as a simple model for endocytosis \cite{vanderhamShallowFullWrapping2025}. We compared two distinct configurations: a spherical vesicle of radius $R_{\text{ves}} \approx 29 \sigma$ (12500 particles) and a planar periodic membrane patch of size $L=50 \sigma$ (2500 particles). Membrane model parameters are set to ensure a soft and diffusive lipid bilayer: $k_{\text{tilt}}=12$, $k_{\text{splay}}=1$, $\zeta=5.0$, $w_{\text{c}}=2.0 \sigma$ and a cutoff radius $r_{\text{c}}=2.5 \sigma$.
The colloid is modeled as a spherical particle which interacts with the membrane through a 12-6 Lennard-Jones potential with a minimum at $\sigma_{\text{mix}} = (\sigma + \sigma_{\text{coll}})/2$ and a varying attraction strength $\varepsilon_{\text{coll}}$. The potential was truncated at a cutoff $r_{\text{coll}} = \sigma_{\text{mix}} + 2\sigma$. Simulations were performed using Langevin dynamics as detailed in Sec.~\ref{sec:methods}. For the planar system, we utilized a barostat to maintain zero lateral tension, ensuring that the wrapping process was governed primarily by the interplay of adhesion and bending elasticity rather than membrane stretching. In the initial configuration of the simulation we placed the colloidal particle at a distance $d=r_{\text{coll}}+2\sigma$ from an equilibrated membrane and gave it a small velocity directed towards the membrane. Each system was simulated for $5\times 10^{5}$ steps with time step $\Delta t = 0.01 \tau_{\text{LJ}}$, which allows it to reach a steady state for the process.

\begin{figure*}
    \centering
    \includegraphics[width=\textwidth]{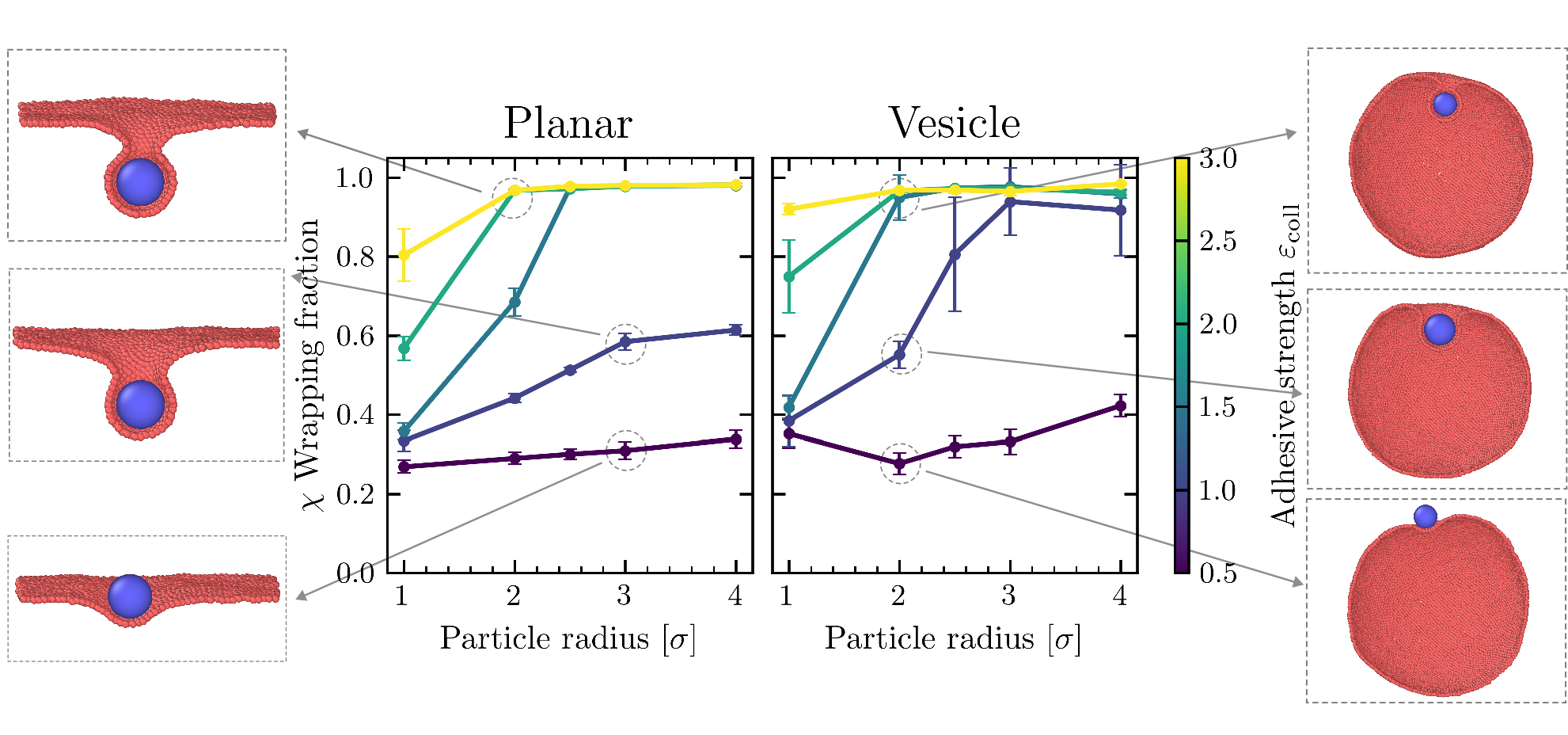}
    \caption{Wrapping fraction~$\chi$ of the colloidal particle as a function of attraction strength $\varepsilon_{\text{coll}}$ (indicated in the colorbar) and particle radius $r_{\text{coll}}$. (left) Interaction with a planar periodic membrane patch ($L=50 \sigma$). (right) Interaction with a spherical vesicle ($R_{\text{ves}} \approx 29 \sigma$). In both plots, error bars represent the standard deviation of the last 15 simulation frames.}
    \label{fig:wrapping_stable}
\end{figure*}

We determined the particle wrapping fraction~$\chi$ by identifying the membrane beads which are within the interaction radius and fitting a circular arc to their coordinates on the midplane. This geometric fit allows for a robust estimation of the extent of the contact zone. The data presented in Fig.~\ref{fig:wrapping_stable} represent averages over the final 15 frames of each simulation.

We observed consistent behavior across both the vesicle and the planar patch geometries: the wrapping fraction increases monotonically with the interaction strength $\varepsilon_{\text{coll}}$ and decreases with the colloidal radius $R_{\text{coll}}$. While a larger radius increases the possible contact area, it also imposes a higher energetic cost due to the bending required to follow the particle's curvature. These results are in qualitative agreement with theoretical predictions for the wrapping of rigid spheres by fluid membranes \cite{Deserno2004}, where the transition from partial to full wrapping is determined by the balance between adhesive energy gain and the bending energy penalty.

\subsection{Volume control protocol}
To demonstrate the model's extensibility to realistic experimental conditions, such as osmotic shock, we implemented a volume control protocol based on the methodology established by \cite{vanhille-camposModellingDynamicsVesicle2021,yuanDynamicShapeTransformations2010}. The protocol entails adding explicit solvent particles inside and outside a vesicle creating an effective pressure. This approach allows us to investigate how the vesicle adapts its morphology under varying pressure differences; a description of the simulation procedure is provided in Supplementary Materials Sec.~\ref{sec:lammps_details}.

To characterize the resulting morphologies, we calculated the reduced volume $\nu$, a dimensionless parameter that measures the deviation of the vesicle from a spherical shape:
\begin{equation}
\nu = \frac{V_{\text{ve}}}{\frac{4}{3} \pi R_{\text{ve}}^3}.
\end{equation}

\begin{figure}[ht]
    \centering
    \includegraphics[width=\columnwidth]{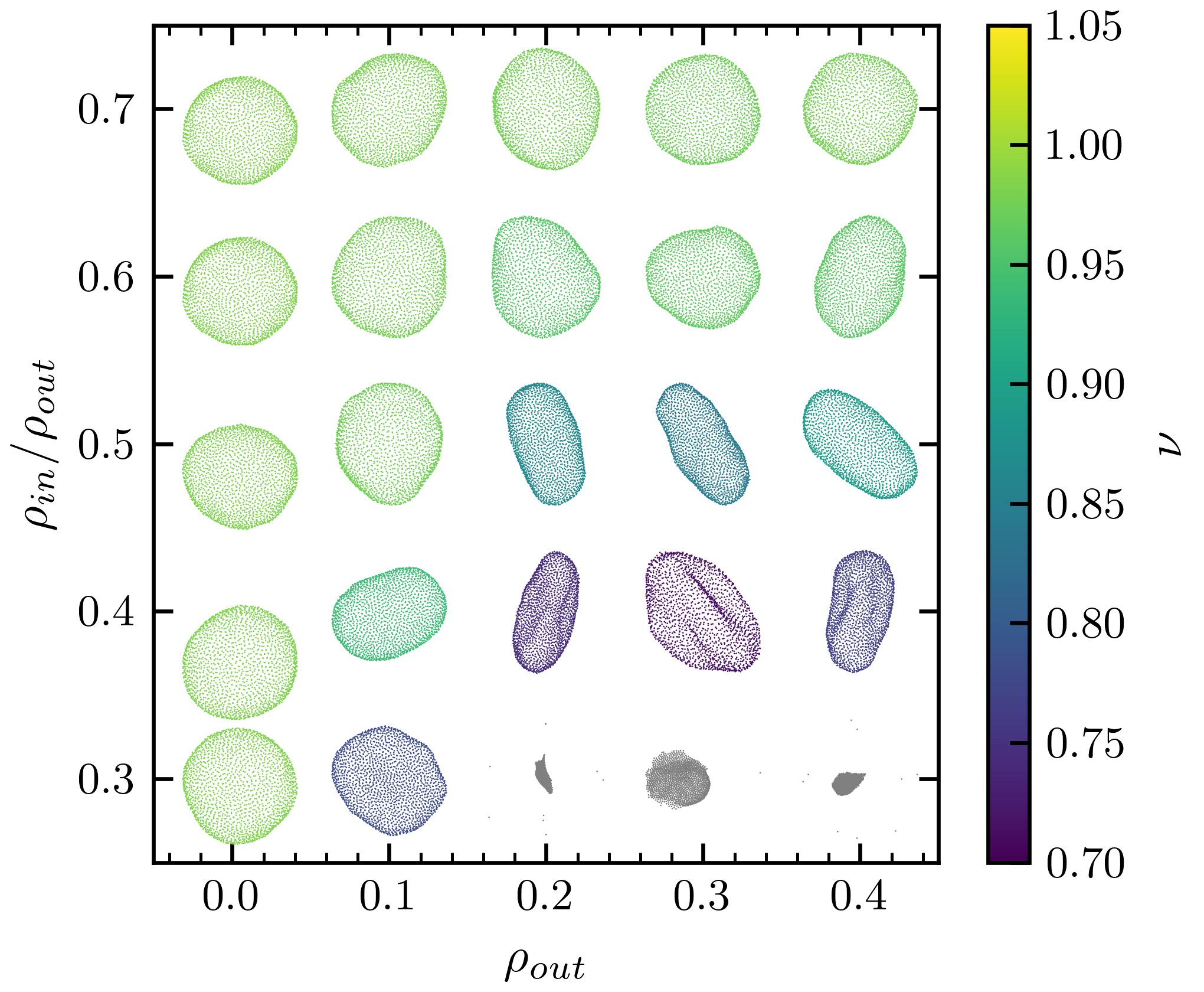}
    \caption{Phase diagram of vesicle morphologies as a function of outer particles density (x-axis) and the internal/external density ratio (y-axis). The vesicles are color-coded according to the reduced volume $\nu$, illustrating the transition from spherical to deflated states. Burst or unstable vesicles are colored in grey.}
    \label{fig:volume_control}
\end{figure}

Here, $R_{\text{ve}}$ represents the equivalent radius of a sphere with the same surface area $A_{\text{ve}}$ as the vesicle, defined as:
\begin{equation}
R_{\text{ve}} = \sqrt{\frac{A_{\text{ve}}}{4\pi}}.
\end{equation}
The relationship between the external density, the internal/external concentration ratio, and the resulting vesicle shape is summarized in Fig.~\ref{fig:volume_control}. The resulting shapes qualitatively agree with numerical solutions of the axisymmetric vesicle shape equation derived from the Helfrich energy~\cite{deulingCurvatureElasticityFluid1976,Seifert1991}.

\subsection{Validity, limitations and future extensions of the model}
Due to its mesoscopic scale, our model can be used to simulate membrane processes at relatively large sizes, relevant for the study of cells and Giant Unilamellar Vesicles (GUVs), with relatively low numbers of particles. Inherently, processes at the scale of our basic unit, the membrane bead of approximately $5\;\mathrm{nm}$ in diameter, cannot be accurately described by our model. For example, while membranes built of our model beads can undergo fission, \textit{e.g.} in cell division, real lipid bilayer membranes will undergo a more detailed process that involves hemifusion of leaflets~\cite{Kozlovsky2003,Mattila2015,Zhao2016}. Likewise, we cannot capture differences in leaflet composition (apart from an induced spontaneous curvature) or relative leaflet motion. We can incorporate multiple lipid domains by including different beads (see Fig.~\ref{fig:panel_systems}C), which could also have different sizes, as set by the correspoinding lipids.

Possible extensions of the model include coarse-grained models of proteins consisting of beads connected by springs. These could represent cytoskelal elements, such as microtubuli and actin filaments, or membrane-associated proteins, as done for example in Helle et al.~\cite{Helle2017} for mitochondrial form factors. Similarly, the model could be used to study the interaction between a membrane and an encapsulated semiflexible polymer, such as DNA.

\section{Conclusion}
We have developed an efficient and pragramatic particle based model for simulating lipid bilayer membranes, based on a conceptually simple additive potential combining isotropic and orientational contributions. We have shown that with this model, we can recover essential membrane properties such as spontaneous vesiculation, the correct scaling of the fluctuation spectrum, and biologically relevant bending and compressibility moduli. The model can be easily extended to include multiple lipid types and spontaneous curvature effects. We can also include an effective osmotic pressure to control the membrane's surface-to-volume ratio. We have demonstrated a possible application of this model to colloidal particle wrapping.

Operating at the mesoscopic level, our model can reach length and timescales relevant for cellular processes, including cell motility, cell growth and cell division. This model cannot represent processes where the one bead-bilayer patch is not a valid approximation anymore. We can further extend the model to include interactions with cytoskeletal elements and membrane-associated proteins, and adhesion between membranes. As such, it complements and bridges between existing molecular-scale models and continuum models.

\section*{Author contributions}
PS and TI conceived the original idea and developed the theoretical framework. PS designed and performed the simulations, analyzed the data, and wrote the initial manuscript. SM and TI acquired funding, supervised, provided critical feedback, and helped shape the final version of the manuscript.

\section*{Acknowledgments}
We thank Felix Frey and Adam Prada for helpful discussions.

This research is supported by the `BaSyC - Building a Synthetic Cell' Gravitation grant (024.003.019) of the Netherlands Ministry of Education, Culture and Science (OCW) and the Netherlands Organisation for Scientific Research (NWO).

\section*{Code and data availability}
The full codebase used for simulations and analysis is openly available on GitLab at \url{https://gitlab.tudelft.nl/idema-group/MesoMem}~\cite{MesoMem}. The data that supports the findings of this article are openly available at~\cite{MesoMemData}.

\clearpage

\onecolumngrid
\appendix

\section{Model description}
\label{sec:modeldescription}


\subsection{Useful math relations}

We first establish the decomposition of a vector $\vec{n}$ into components parallel and perpendicular to the radial unit vector $\hat{r}$:
$$
\vec{n} = n^{\parallel} \hat{r} + n^\perp \hat{r}^\perp = \left[ (\vec{n} \cdot \hat{r})\hat{r}\right] + \left[\vec{n} - (\vec{n} \cdot \hat{r})\hat{r}\right].
$$
Doing so allows us to define the perpendicular projection operator, represented by the matrix $\mathbf{P}_{\vec{v}^{\perp}}$. This matrix acts on a vector $\vec{u}$ to isolate its transverse component relative to a reference vector $\vec{v}$:

\begin{equation} \label{eq:proj_perp}
    \mathbf{P}_{\vec{v}^{\perp}}(\vec{u}) = \vec{u} - \frac{\vec{u}\cdot \vec v}{v^2}\vec{v} = \left[\mathbf{I} - \frac{\vec{v} \otimes \vec{v}}{v^2} \right] \vec{u}.
\end{equation}

We can rewrite the gradient acting on a vector field in terms of the projection operator:
$$
\vec{\nabla} \hat{r} = \frac{\partial \hat{r}}{\partial \vec{r}} =\frac{1}{r} \left(\mathbf{I} - \frac{\vec{r} \otimes \vec{r}}{r^2}\right) = \frac{1}{r} \mathbf{P}_{\hat{r}^{\perp}}.
$$

This operator-based approach can be extended to scalar fields $U$ that depend on the orientation $\hat{r}$:

$$
\vec{\nabla} U = \frac{\partial U}{\partial \vec{r}} = \frac{\partial \hat{r}}{\partial \vec{r}} \cdot \frac{\partial U}{\partial \hat{r} }  = \frac{1}{r} \left(\mathbf{I} - \frac{\vec{r} \otimes \vec{r}}{r^2}\right) \frac{\partial U}{\partial \hat{r}} = \frac{1}{r} \mathbf{P}_{\hat{r}^{\perp}} \left( \frac{\partial U}{\partial \hat{r}} \right).
$$

\begin{figure}[ht]
    \includegraphics[width=0.95\textwidth]{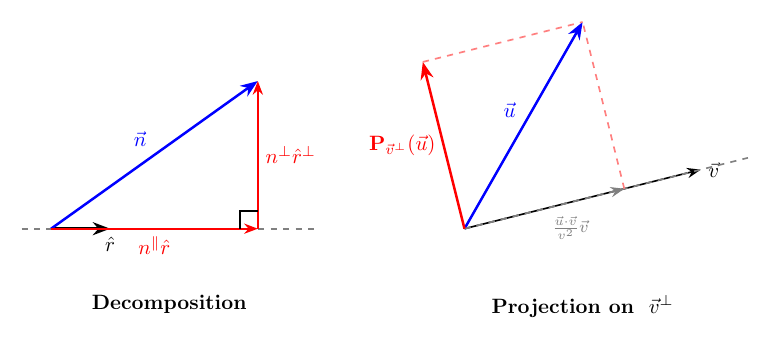}
    \centering
    \label{fig:S0}
\end{figure}

\subsection{Model definition}

Let $\vec{r}_{ij} = \vec{r}_i - \vec{r}_j$ be the separation vector between particle $i$ and particle $j$, with its magnitude defined as $r = |\vec{r}_{ij}|$. The unit vector $\hat{r}$ is defined as the normalized direction pointing from particle $j$ to particle $i$:
$$
\hat{r} = \frac{\vec{r}_i - \vec{r}_j}{r}.
$$

The total energy of our system is given by the sum over the isotropic and the orientational terms, the latter of which are modulated by a distance-dependent weighting function (see Eq.~\ref{modeltotalenergy} in the main text):
$$
H = U_{\text{rep}} + U_{\text{attr}} + w(r)\left[U_{\text{tilt}} + U_{\text{splay}}\right].
$$

\subsubsection{Weighting function}
\paragraph{Force derivation in presence of weighting function}

All orientational potentials in this model are scaled by a radial weighting function $w(r)$. To derive the total force, we apply the chain rule for the derivative with respect to $\vec{r}$:

\begin{align}\label{eq:weight_f_eq}
    \vec{F}_{\text{tot}} &= 
    - \nabla_{\vec{r}} \big[U(\hat{r}) w(r)\big] = - w(r)\frac{\partial U(\hat{r})}{\partial \hat{r}} \frac{\partial \hat{r}}{\partial \vec{r}} - U(\hat{r})\frac{\partial w(r)}{\partial r} \frac{\partial r}{\partial \vec{r}} \nonumber\\
    &=w(r) \vec{F} - U(\hat{r})w^\prime(r) \hat{r}.
\end{align}

The first term in equation~\eqref{eq:weight_f_eq} is the force due to the potential $U(\hat{r})$ weighted by $w(r)$, without changing direction. The second term is proportional to the interaction energy and to the derivative of the weighting function.

\paragraph{Torque derivation in presence of weighting function}

The torques are simpler because for them the weighting function is just a rescaling:
$$
\vec{\tau}_{\text{tot}} = 
- \nabla_{\hat{n}} \big[U(\hat{r}) w(r)\big] = - w(r)\frac{\partial U(\hat{r})}{\partial \hat{n}} = - w(r) \vec{\tau}.
$$

\paragraph{Functional form of the weighting function $w(r)$}
We have chosen the functional form for our weighting function $w(r)$ to interpolate smoothly between $1$ at $r=0$ and $0$ at the cutoff distance $r = w_\mathrm{c}$. The derivative of $w(r)$ vanishes at both $r=0$ and $r=w_\mathrm{c}$, ensuring no jump in force or torque as a particle comes within range.
\begin{equation}
\label{supp-tiltsplaycutoff}
w(r) =
\begin{cases}
\exp \left( \dfrac{(r/r_\text{ga})^2}{(r/w_c)^n - 1} \right), & r \leq w_\mathrm{c}, \\[1ex]
0, & r \geq w_\mathrm{c},
\end{cases}
\end{equation}
typically with $n=4$ and $r_\text{ga}=1.5\sigma$. The freely adjustable parameter $w_\mathrm{c}$ is used to set the effective range of orientational interactions.

\begin{figure}[ht]
    \includegraphics[width=0.7\textwidth]{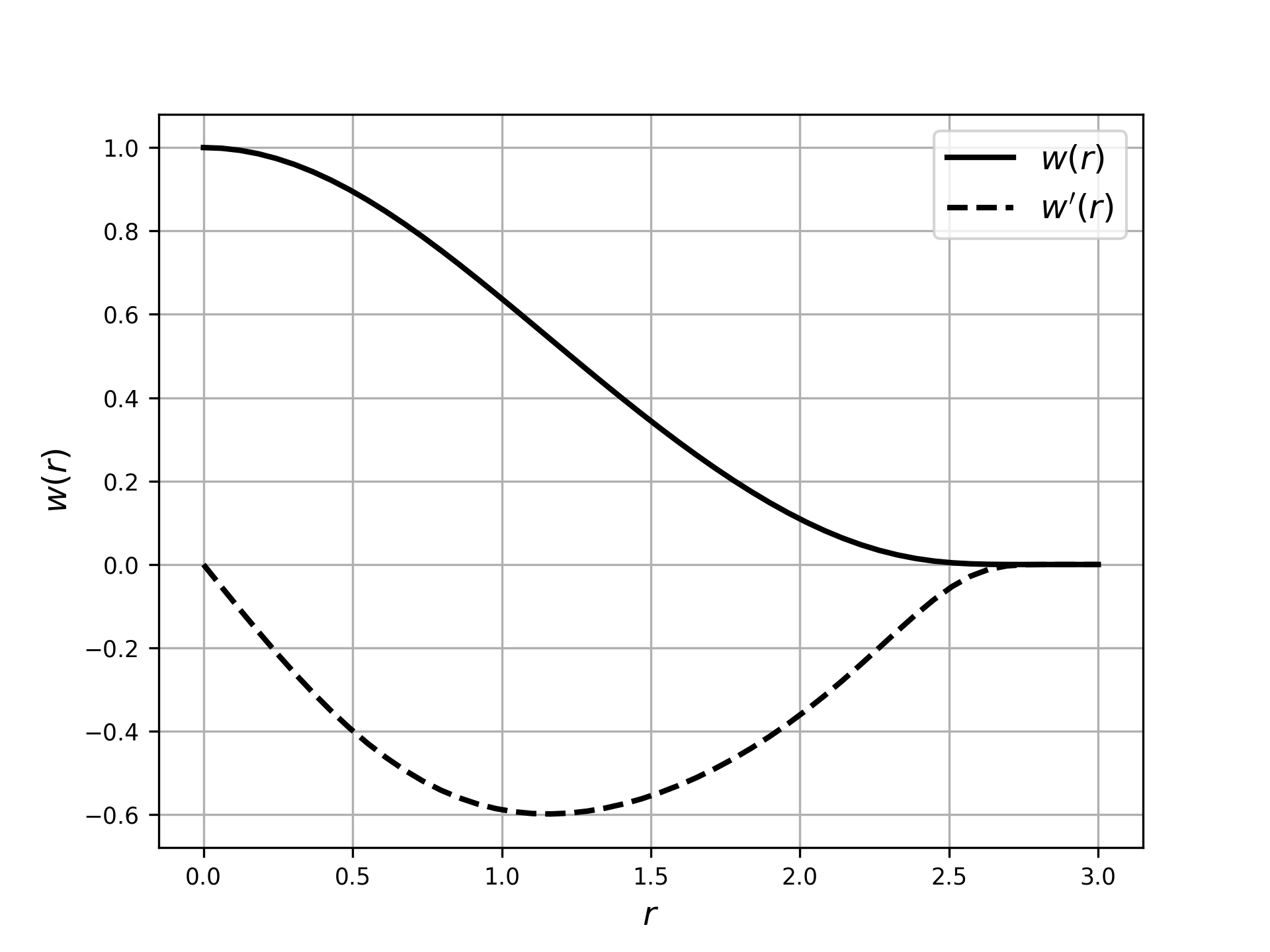}
    \centering
    \caption{Plot of weighting function $w(r)$ used in the membrane model.}
    \label{fig:S1}
\end{figure}

\clearpage
\subsection{Forces}
\subsubsection{Isotropic potential}
For the forces due to the isotropic potential, we obtain:
\begin{align*}
\vec{F} &= - \nabla U(r) = - \frac{\partial U}{\partial \vec{r}} = - \frac{\partial U}{\partial r} \frac{\partial r}{\partial \vec{r}} = - \frac{\partial U}{\partial r} \frac{\vec{r}}{r} = - \frac{\partial U}{\partial r} \hat{r},\\
U_{\text{rep}}(r) &= \epsilon \left[ \left( \frac{r_{\min}}{r} \right)^{4} -2 \left( \frac{r_{\min}}{r} \right)^{2} \right], 
\qquad r \leq r_{\min},\\
\vec{F}_{\text{rep}}(r) &=  \frac{4 \epsilon}{r} \left[ \left( \frac{r_{\min}}{r} \right)^{4} - \left( \frac{r_{\min}}{r} \right)^{2} \right] \hat{r}, 
\qquad r \leq r_{\min},\\
U_{\text{attr}}(r) &=
\begin{cases}
- \epsilon \cos^{2 \zeta}\left( \dfrac{\pi (r - r_{\min})}{2 \,(r_c - r_{\min})} \right), & r_{\min} < r < r_c, \\[1ex]
0, & r \geq r_c,
\end{cases} \\
\vec{F}_{\text{attr}}(r) &= \epsilon \frac{\pi}{r_c - r_{\min}} \zeta \cos^{2 \zeta - 1} \left(\dfrac{\pi (r - r_{\min})}{2 \,(r_c - r_{\min})} \right) \sin \left(\dfrac{\pi (r - r_{\min})}{2 \,(r_c - r_{\min})} \right) \hat{r}
\end{align*}

\subsubsection{Tilt}
\begin{align*}
U_{\text{tilt}} &= \frac{k_{\text{tilt}}}{2} \left[(\hat{n}_i \cdot \hat{r})^2 + (\hat{n}_j \cdot \hat{r})^2 \right],\\
\vec{\nabla} U &= \frac{\partial U}{\partial \vec{r}} = \frac{\partial U}{\partial \hat{r} } \frac{\partial \hat{r}}{\partial \vec{r}} = \frac{1}{r} \frac{\partial U}{\partial \hat{r}} \left(\mathbf{I} - \frac{\vec{r} \otimes \vec{r}}{r^2}\right) = \frac{1}{r} \frac{\partial U}{\partial \hat{r}}  \mathbf{P}_{\vec{r}^{\perp}},\\
\vec{F} &= - \nabla U_{\text{tilt}} = - \frac{\partial }{\partial \vec{r}} U_{\text{tilt}}(\hat{r}\cdot \hat{n}_i,\hat{r}\cdot \hat{n}_j ) = - \frac{\partial U_{\text{tilt}}}{\partial \hat{r}} \frac{\partial \hat{r}}{\partial \vec{r}} =  -  \frac{\partial U_{\text{tilt}}}{\partial \hat{r}}  \frac{1}{r} \mathbf{P}_{\vec{r}^{\perp}},
\end{align*}
using Eq.~\eqref{eq:proj_perp}

\paragraph{Force on particle $i$}
\begin{align*}
\vec{F_i} &= - \frac{k_\text{tilt}}{2}\left[ 2(\hat{n}_i \cdot \hat{r}) \hat{n}_i + 2(\hat{n}_j \cdot \hat{r}) \hat{n}_j\right] \frac{\partial \hat{r}} {\partial \vec{r}} \\
&= - \frac{k_\text{tilt}}{2} \left[2(\hat{n}_i \cdot \hat{r}) \frac{1}{r} \left( \hat{n}_i - (\hat{n}_i\cdot \hat{r})\hat{r}\right) +2 (\hat{n}_j \cdot \hat{r}) \frac{1}{r} \left( \hat{n}_j - (\hat{n}_j\cdot \hat{r})\hat{r}\right) \right]\\
&= -\frac{k_\text{tilt}}{2} \left[ 2  \frac{(\hat{n}_i \cdot \hat{r})}{r} \vec{n}_i^\perp +2  \frac{(\hat{n}_j \cdot \hat{r})}{r}\vec{n}_j^\perp \right],
\end{align*}
where $\vec{n}_i^\perp = \hat{n}_i-(\hat{n}_i\cdot \hat{r})\hat{r}$ is the projection of $\hat{n}_i$ onto the plane perpendicular to $\hat{r}$.


\paragraph{Force on particle $j$}
Substituting the flipped unit vector into the expression for $\vec{F}_i$, we find:
\begin{align*}
\vec{F}_j &= +\frac{k_\text{tilt}}{r} \left[ (\hat{n}_i \cdot \hat{r})\vec{n}_i^\perp + (\hat{n}_j \cdot \hat{r})\vec{n}_j^\perp \right].
\end{align*}
This yields the relation $\vec{F}_j = -\vec{F}_i$, confirming that the internal forces of the pair interaction satisfy Newton's third law and conserve the total linear momentum of the system.



\subsubsection{Splay}
\begin{equation}
\label{supp-Usplay}
U_{\text{splay}} = \frac{k_{\text{splay}}}{2}\left(\hat{n}_i \cdot \hat{n}_j - 1\right)^2.
\end{equation}
The splay term is not generating any force onto the interacting pair.

\subsection{Torques}
Following the derivation from Allen~\cite{allenExpressionsForcesTorques2006}, the torques can be expressed as:

\begin{align*}
\vec{\tau}_i &= \frac{\partial U}{\partial (\hat{n}_i \cdot \hat{r})} (\hat{r} \times \hat{n}_i) - \frac{\partial U}{\partial (\hat{n}_i \cdot \hat{n}_j)} (\hat{n}_i \times \hat{n}_j) \\
\vec{\tau}_j &= \frac{\partial U}{\partial (\hat{n}_j \cdot \hat{r})} (\hat{r} \times \hat{n}_j) + \frac{\partial U}{\partial (\hat{n}_i \cdot \hat{n}_j)} (\hat{n}_i \times \hat{n}_j)
\end{align*}

\subsubsection{Isotropic potential}
The isotropic potential term is not generating any torque onto the interacting pair.

\subsubsection{Tilt}

\begin{align*}
\vec{\tau_i} &= \frac{k_{\text{tilt}}}{2} 2(\hat{n}_i \cdot \hat{r}) (\hat{r} \times \hat{n}_i) = \frac{k_{\text{tilt}}}{2} 2(\hat{n}_i \cdot \hat{r}) (\hat{r} \times \hat{n}_i), \\
\vec{\tau_j} &= \frac{k_{\text{tilt}}}{2} 2(\hat{n}_j \cdot -\hat{r}) (-\hat{r} \times \hat{n}_j) = \frac{k_{\text{tilt}}}{2} 2(\hat{n}_j \cdot \hat{r}) (\hat{r} \times \hat{n}_j).
\end{align*}

\subsubsection{Splay}
\begin{align*}
\tau_i &= -\frac{k_{\text{splay}}}{2} \, 2(\hat{n}_i \cdot \hat{n}_j  - 1) (\hat{n}_i \times \hat{n}_j) = - k_{\text{splay}} (\hat{n}_i \cdot \hat{n}_j  - 1) (\hat{n}_i \times \hat{n}_j), \\
\tau_j &= +  \frac{k_{\text{splay}}}{2} \, 2 (\hat{n}_i \cdot \hat{n}_j - 1) (\hat{n}_i \times \hat{n}_j) = +  k_{\text{splay}} (\hat{n}_i \cdot \hat{n}_j  - 1) (\hat{n}_i \times \hat{n}_j).
\end{align*}

\clearpage
\subsection{General case with spontaneous curvature}\label{sec:curvature}
\begin{figure}[ht]
    \includegraphics[width=0.45\textwidth]{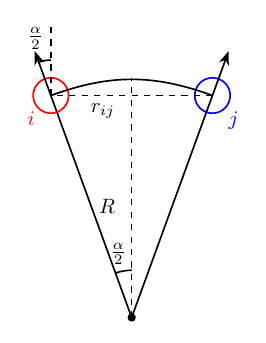}
    \centering
    \caption{Schematic for spontaneous curvature geometry.}
    \label{fig:S2}
\end{figure}

If we include a non-zero spontaneous curvature as described in Sec.~\ref{sec:spont_curv}, the tilt and splay energies are modified to:

\begin{align*}
U_{\text{tilt}} &= \frac{k_{\text{tilt}}}{2} \left[ (\hat{n}_i \cdot \hat{r} + \sin(\alpha/2))^2 + (\hat{n}_j \cdot \hat{r} - \sin(\alpha/2))^2 \right] \\
&= \frac{k_{\text{tilt}}}{2} \left[ \left(\hat{n}_i \cdot \hat{r} + \frac{1}{2} r C_0 \right)^2 + \left(\hat{n}_j \cdot \hat{r} - \frac{1}{2} r C_0\right)^2 \right], \\
U_{\text{splay}} &= \frac{k_{\text{splay}}}{2} \left(\hat{n}_i \cdot\hat{n}_j - \cos \alpha\right)^2 \\ 
&= \frac{k_{\text{splay}}}{2} \left[\hat{n}_i \cdot\hat{n}_j -1 + 2\left(\frac{1}{2}C_0 r\right)^2 \right]^2. \\
\end{align*}

The preferred angle $\alpha$ is coupled to the distance $r$ between the particles according to $\sin \frac{\alpha}{2} = \frac{1}{2} r C_0$, where $C_0$ is the spontaneous curvature. In the limit where the spontaneous curvature vanishes ($C_0 \to 0$) the simplified form is recovered.

\subsubsection{Tilt}
Force on particle $i$:
\begin{align*}
\vec{F}_i &= -\frac{k_{\text{tilt}}}{r} \left[ (\hat{n}_i \cdot \hat{r} + \frac{1}{2} r C_0) (\hat{n}_i - (\hat{n}_i \cdot \hat{r})\hat{r}) + (\hat{n}_j \cdot \hat{r} - \frac{1}{2} r C_0) (\hat{n}_j - (\hat{n}_j \cdot \hat{r})\hat{r}) \right] \\
&= -\frac{k_{\text{tilt}}}{r} \left[ (\hat{n}_i \cdot \hat{r} + \frac{1}{2} r C_0)\vec{n}_i^\perp + (\hat{n}_j \cdot \hat{r} - \frac{1}{2} r C_0) \vec{n}_j^\perp \right].
\end{align*}
Because the preferred angle is coupled to the separation through $\sin(\alpha/2) = \frac{1}{2} r C_0$, the tilt energy now also depends explicitly on $r$ and therefore contributes a radial force, which is absent for $C_0 = 0$:
$$
\vec{F}_i^{\,\text{rad}} = \frac{k_{\text{tilt}} C_0}{2} \left[ \left(\hat{n}_j \cdot \hat{r} - \frac{1}{2} r C_0\right) - \left(\hat{n}_i \cdot \hat{r} + \frac{1}{2} r C_0\right) \right] \hat{r}.
$$
Torque on particle $i$:
$$
\vec{\tau}_i = k_{\text{tilt}} \left(\hat{n}_i \cdot \hat{r} + \frac{1}{2} r C_0\right)(\hat{r} \times \hat{n}_i).
$$

\subsubsection{Splay}
Unlike the case of zero spontaneous curvature, the generalized splay term results in a force acting onto the interacting particles:
\begin{align*}
    \vec{F}_i &= - k_{\text{splay}} C_0^2 r \left[ \hat{n}_i \cdot \hat{n}_j - 1 + 2 \left(\frac{1}{2} r C_0\right)^2 \right] \hat{r}\\
\end{align*}
Torque on particle $i$:
\begin{align*}
\vec{\tau}_i &= -k_{\text{splay}} \left[ \hat{n}_i \cdot \hat{n}_j - 1  + \frac{1}{2} (r C_0)^2 \right] (\hat{n}_i \times \hat{n}_j).
\end{align*}

\clearpage
\section{System preparation} \label{sec:init}
\subsection{Planar membrane initial configuration}\label{sec:plane}
As already mentioned in Sec.~\ref{sec:stabilityconditions} of the main text, we have used a two-dimensional hexagonal packing as the template for the initial configuration. The distance between adjacent hexagonal cells is $0.8 \sigma$. For these simulations we use periodic boundary conditions, so edge particles need to be close enough to the box edge to interact with their periodic neighbors to conserve plane continuity.

\subsection{Spherical vesicle initial configuration}\label{sec:vesicle}

To initialize the vesicle, we used the icosphere package \cite{icosphere} to generate a triangular mesh of a sphere. This method takes an icosahedron as a starting point, and then iteratively subdivides its triangles, creating a mesh of the sphere. We then place single particles with outward orientation at the barycenter of each triangle. 
An important final step is to rescale the neighbors distance to around $0.8 \sigma$.

\subsection{Lipid tube initial configuration}\label{sec:tube}
To generate a lipid tube, we extend the setup used for the planar membrane by mapping the hexagonal lattice onto a cylindrical geometry. The initial configuration is defined by a radius $R$ and a total length $L$ along the longitudinal axis, typically aligned with the $z$-direction of the simulation box. We also check that beads at the top and bottom boundaries can interact smoothly under periodic boundary conditions. Similar to the planar case, the equilibrium distance between nearest neighbors is scaled to approximately $0.8 \sigma$. 

\subsection{Volume control configuration}\label{sec:volume_control}
To simulate an osmotic shock and achieve precise control over the vesicle volume, we introduce explicit solvent particles into the system. These particles generate an effective osmotic pressure across the lipid bilayer, which can be finely tuned by varying the particle number density inside ($\rho_\text{in}$) and outside ($\rho_\text{out}$) the vesicle. 

The solvent is composed of simple point-like particles modeled using a standard repulsive branch of Lennard-Jones (12-6) potential truncated at its minimum ($r_\text{c} = 2^{1/6}\sigma$) to ensure excluded volume interactions. The solvent particles interact with the membrane beads through the same truncated LJ potential.

\subsection{Details on MD Simulations}\label{sec:lammps_details}

To stabilize this initial configuration, we performed an energy minimization step in LAMMPS using the following command:

\begin{verbatim}
    minimize 0.0 1.0e-8 1000 100000
\end{verbatim}

Following the minimization procedure it is possible to start a NVT production  run.  

The molecular dynamics (MD) simulations were performed using the following system specifications and software environment:

\begin{description}
    \item[Software:] \texttt{LAMMPS} (version 29 Aug 2024)
    \item[Operating System:] Linux ``Ubuntu 24.04.3 LTS''
    \item[Compiler \& Parallelization:] \texttt{GNU C++ 11.4.0} with \texttt{Open MPI v4.1.6}
    \item[Installed optional LAMMPS packages:] \texttt{DIPOLE}, \texttt{MOLECULE}, \texttt{EXTRA-FIX}
\end{description}

\clearpage
\section{Simulation snapshots}
\label{sec:snapshots}

\subsection{Self assembly and vesiculation}

\begin{figure}[ht]
    \includegraphics[width=0.95\textwidth]{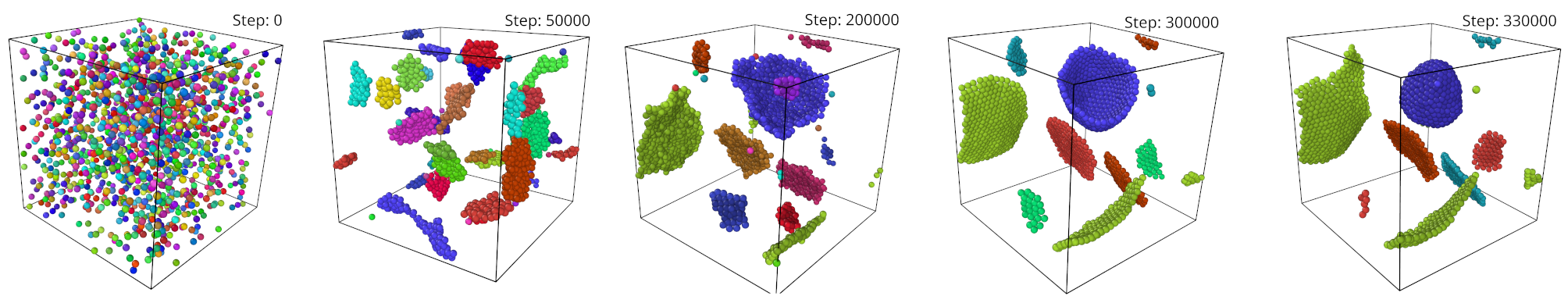}
    \centering
    \caption{Representative snapshots at $t=500\,\tau_{\text{LJ}}$ showing the effect of the tilt modulus $k_{\text{tilt}}$ on self-assembly. For low $k_{\text{tilt}}$ the system forms compact three-dimensional aggregates, while for higher $k_{\text{tilt}}$ extended planar patches appear. Parameters: $k_{\text{splay}}=1$, $\zeta=5.0$, $w_{\text{c}}=2.0 \sigma$ and a cutoff radius $r_{\text{c}}=2.5 \sigma$.}
    \label{fig:S3}
\end{figure}

\begin{figure}[ht]
    \includegraphics[width=0.95\textwidth]{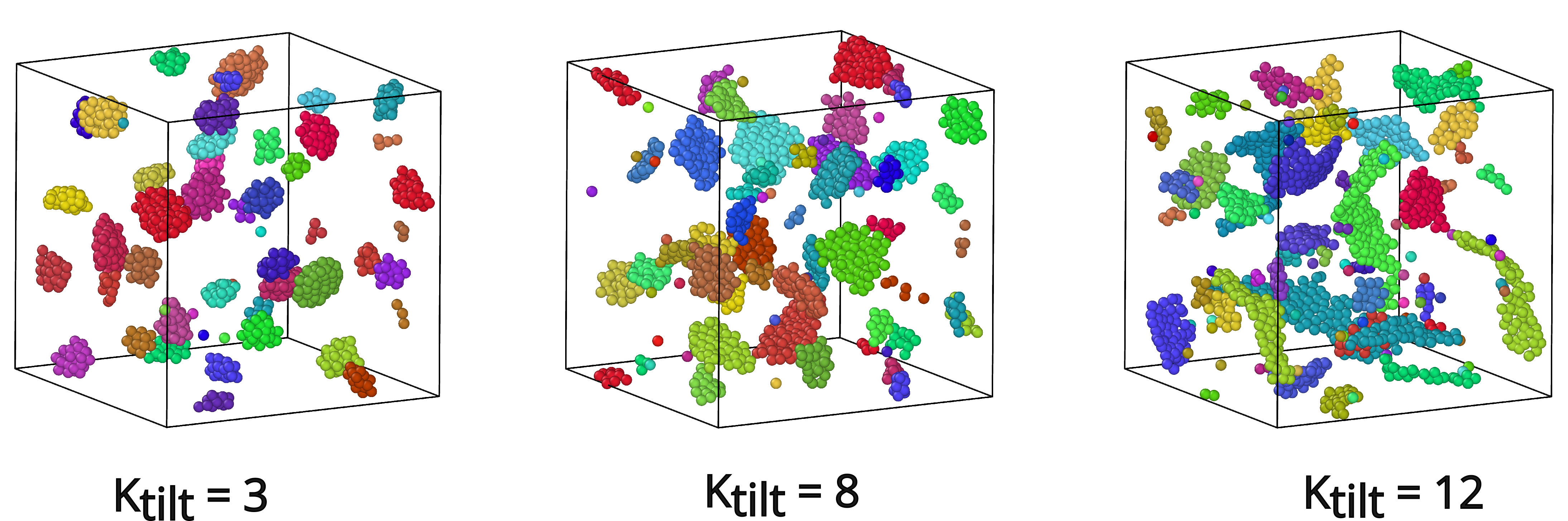}
    \centering
    \caption{Temporal snapshots of self assembly simulations at $t=500\,\tau_{\text{LJ}}$ across a range of tilt moduli. Parameters: $k_{\text{splay}}=1$, $\zeta=5.0$, $w_{\text{c}}=2.0 \sigma$ and a cutoff radius $r_{\text{c}}=2.5 \sigma$.}
    \label{fig:S4}
\end{figure}


\clearpage
\section{Stability exploration}
\begin{figure}[ht]
    \includegraphics[width=0.85\textwidth]{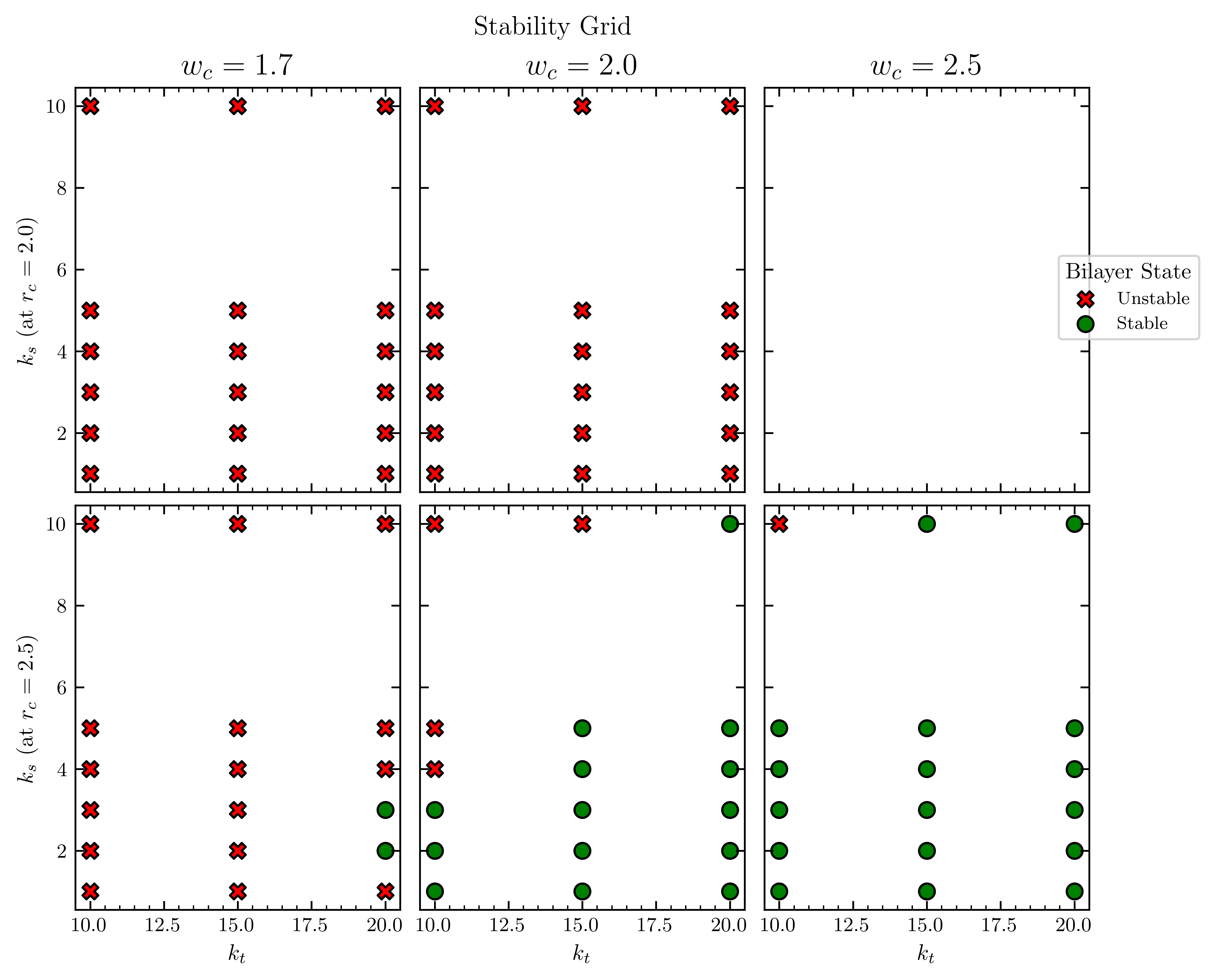}
    \centering
    \caption{Stability plot for different parameters set of the model.}
    \label{fig:S5}
\end{figure}

In Fig.~\ref{fig:S5} we represent the stability plot for different parameters set of the model.

\section{Line tension measurement}
Here we provided the additional figure supporting Sec.~\ref{sec:linetension}:
\begin{figure}[ht]
    \includegraphics[width=0.95\textwidth]{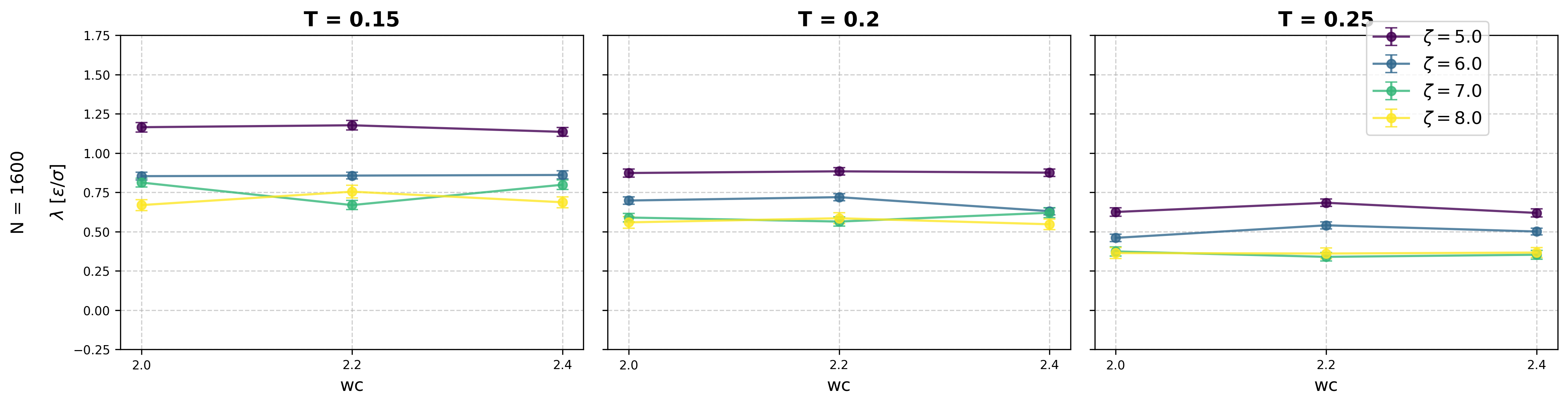}
    \centering
    \caption{Edge tension $\lambda$ as a function of model parameters. Error bars represent the standard error of the mean.}
    \label{fig:tension}
\end{figure}

\section{Tube radius measurement protocol}
\label{sec:tuberadiusmeasurement}
We adopt the protocol of Ref.~\cite{munozbasagoitiTutorialMesoscaleComputer2025}.  A cylindrical lipid tube is initialized and run in two successive phases: Phase~1 equilibrates the tube at zero applied tension to reach its natural radius; Phase~2 applies a target tension via a one-dimensional barostat acting on the axial ($x$) dimension.  The instantaneous radius is computed as the mean radial distance of particles from the tube axis, averaged over the equilibrated portion of Phase~2.  The membrane tension $\gamma$ is extracted from the axial component of the stress tensor via the force balance on the tube end-cap, $\gamma = -P_{xx}\,L_y L_z\,(2\pi R)^{-1}$, where $P_{xx}$ is obtained from the LAMMPS \texttt{compute pressure}.

\begin{figure}[ht]
    \includegraphics[width=0.95\textwidth]{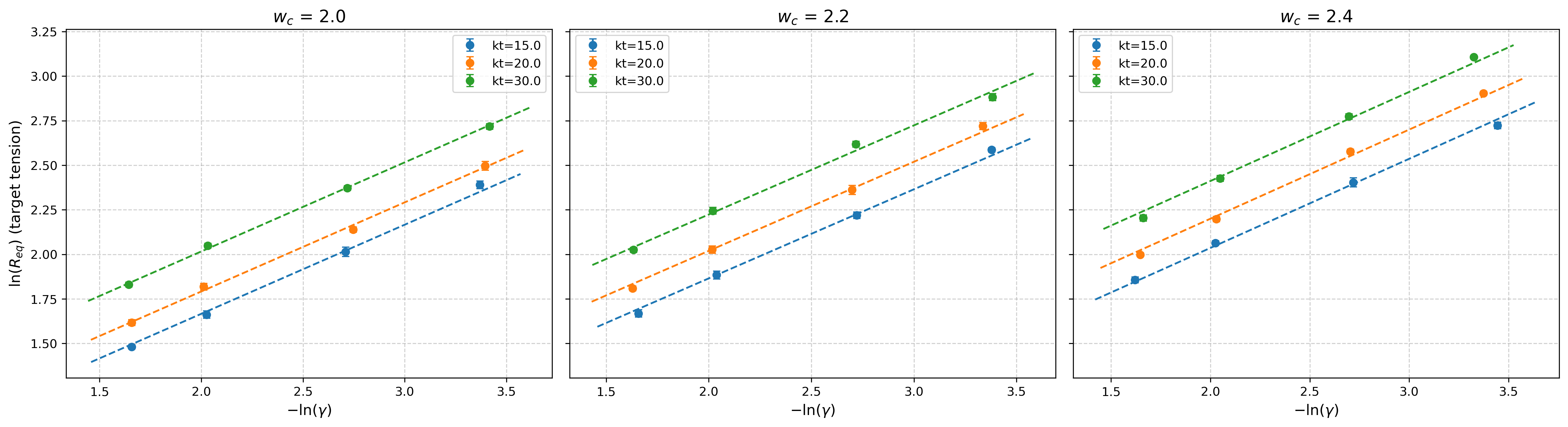}
    \centering
    \caption{Log-log relation between equilibrium tube radius $R_{eq}$ and measured membrane tension $\gamma$, for three values of the potential range $w_c$ (subpanels) and tilt stiffness $k_t$ (colours).  Dashed lines are power-law fits with the exponent fixed to the theoretical value $\alpha = 0.5$. Error bars denote the temporal standard deviation of $R$ over the equilibrated portion of the simulation.}
    \label{fig:tuberadius}
\end{figure}

\section{Computational benchmark}

To evaluate the parallel scalability of the simulation framework, benchmarking was performed on a planar square lattice membrane. The test system utilized an NVE ensemble under periodic boundary conditions. 

We investigated the scaling behavior by sweep-sampling both the total particle count \\($N \in \{400, 2500, 10k, 100k, 500k\}$) and the number of MPI tasks (CPU cores). Performance was quantified using the speedup $S_p = T_1/T_p$, where $T_1$ represents the execution time of the serial (1 CPU core) case. 
All benchmark were performed on a Linux workstation equipped with an AMD Ryzen Threadripper PRO 7985WX CPU (64 cores, 2 threads/core, 3.00 GHz) and 128 GB of RAM, using the LAMMPS configuration specified in Sec.~\ref{sec:lammps_details}.

\begin{figure}[ht]
    \centering
    \begin{subfigure}{0.45\textwidth}
        \centering
        \includegraphics[width=\linewidth]{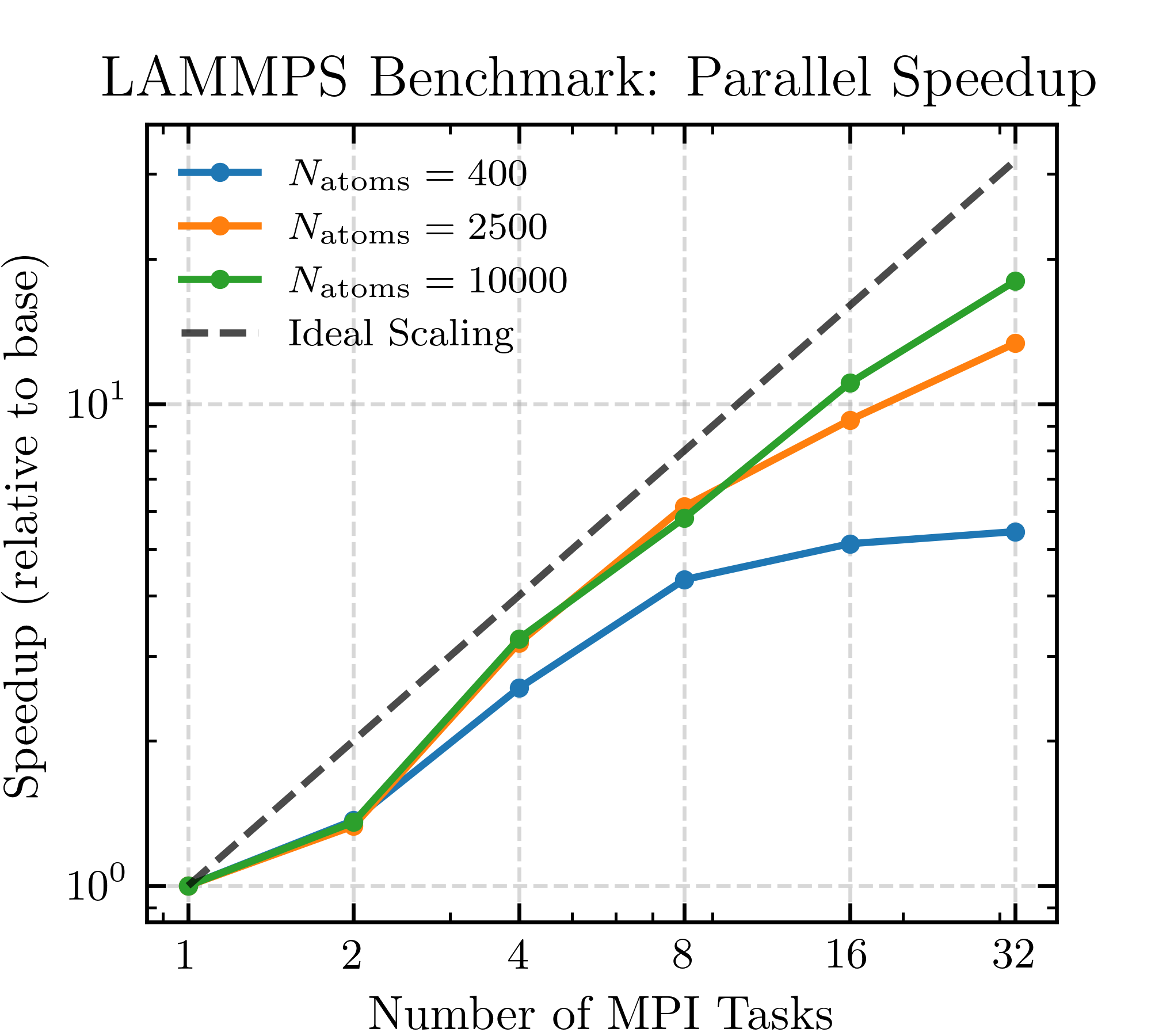}
        \caption{Planar system}
        \label{fig:S8a}
    \end{subfigure}
    \hfill
    \begin{subfigure}{0.45\textwidth}
        \centering
        \includegraphics[width=\linewidth]{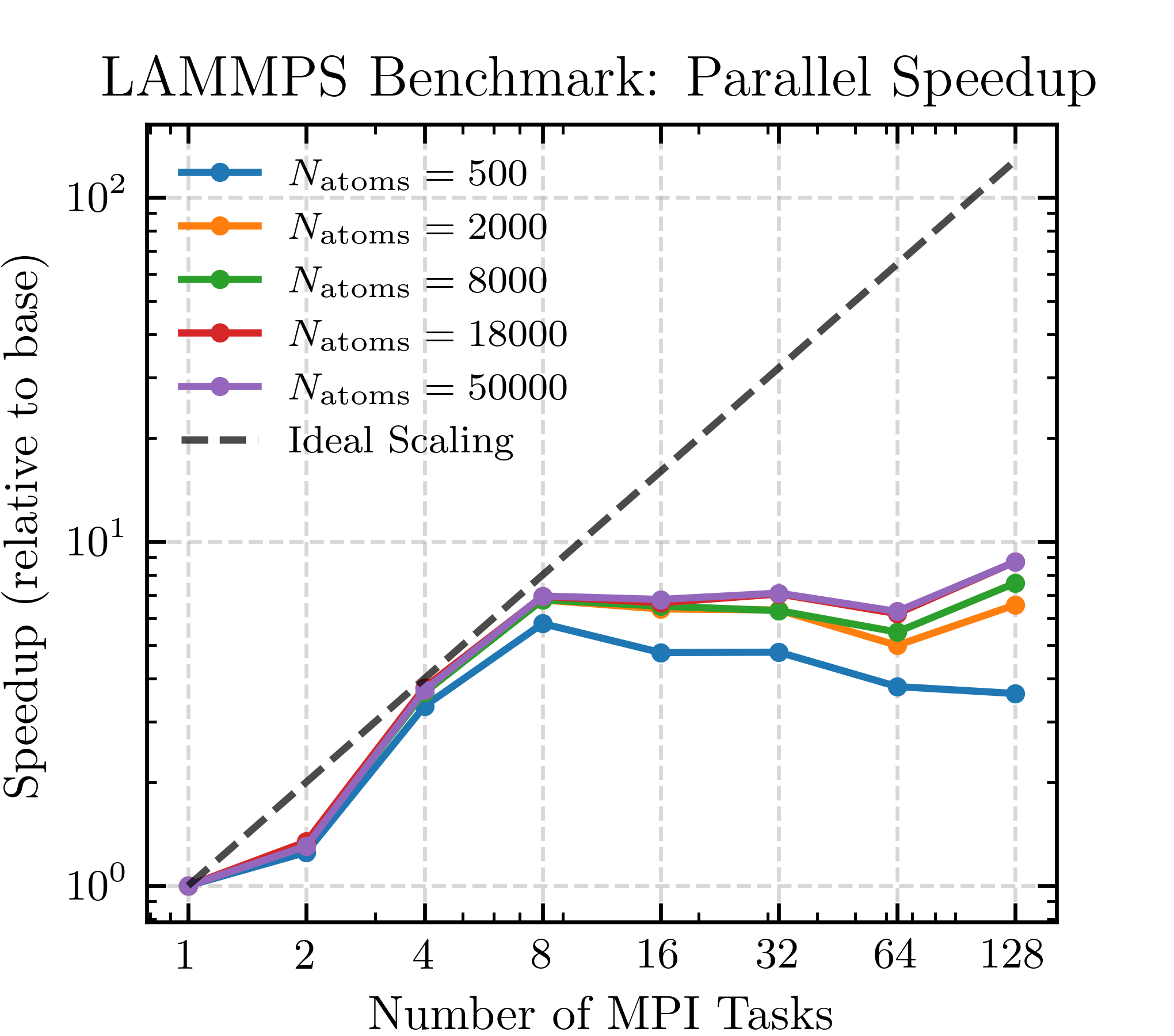}
        \caption{Spherical vesicle}
        \label{fig:S8b}
    \end{subfigure}
    \caption{Parallel speedup as a function of MPI tasks for varying system sizes. The dashed line represents ideal linear scaling.}
    \label{fig:S8}
\end{figure}

Figure~\ref{fig:S8} illustrates the parallel speedup relative to the base configuration for various system sizes. 

\section{GPU implementation}
We ported MesoMem to the Kokkos framework to leverage GPU parallelization for large-scale molecular dynamics simulations. This implementation drastically improves performance for massive systems containing complex components, such as vesicle-polymer complexes or explicit solvent particles. Full implementation details and benchmarks are available in the GPU section of the MesoMem GitLab repository~\cite{MesoMem}.

\newpage


%

\end{document}